\documentclass[useAMS,usenatbib]{mnras}
\usepackage{graphicx,amsmath}
\usepackage{amssymb}
\usepackage{natbib}
\usepackage{pdflscape}
\usepackage{hyperref}
\title[The EBL from FIR to TeV $\gamma$-rays]{\huge
New synthesis models of consistent extragalactic background light over cosmic time
}
\author[Khaire \& Srianand]
{
\parbox{\textwidth}{
Vikram Khaire$^{1, 2}$\thanks{E-mail:vkhaire@physics.ucsb.edu} and Raghunathan Srianand$^{3}$
} 
\vspace*{10pt}\\
$^{1}$Department of Physics, University of California, Santa Barbara, CA 93106-9530, USA\\
$^{2}$National Centre for Radio Astrophysics, Tata Institute of Fundamental Research, Pune 411007, India\\
$^{3}$Inter-University Centre for Astronomy and Astrophysics, Post Bag 4, Pune 411007, India\\  
}   

\begin{document}

\defcitealias{Khaire15ebl}{KS15b}
\defcitealias{Khaire15puc}{KS15a}
\defcitealias{HM12}{HM12}
\defcitealias{Madau15}{MH15}
\defcitealias{FG09}{FG09}

\date{}

\pagerange{\pageref{firstpage}--\pageref{lastpage}} \pubyear{2015}
\maketitle

\label{firstpage}

\vspace{20 mm}

\newcommand{\hi}{H$\,\,{\rm {\scriptstyle I}}$}
\newcommand{\hei}{He$\,\,{\rm {\scriptstyle I}}$}
\newcommand{\heii}{He$\,\,{\rm {\scriptstyle II}}$}
\newcommand{\nhi}{$N_{\rm H\,{\rm {\scriptstyle I}}}$}
\newcommand{\nhei}{$N_{\rm He\,{\rm {\scriptstyle I}}}$}
\newcommand{\nheii}{$N_{\rm He\,{\rm {\scriptstyle II}}}$}
\newcommand{\hii}{H$\,\,{\rm {\scriptstyle II}}$}
\newcommand{\heiii}{He$\,\,{\rm {\scriptstyle III}}$}
\newcommand{\thi}{$\tau_{\rm \alpha}^{\rm H\,{\scriptscriptstyle I}}$}
\newcommand{\theii}{$\tau_{\rm \alpha}^{\rm He\,{\scriptscriptstyle II}}$}
\newcommand{\fesc}{$f_{\rm esc}$~}
\newcommand{\fescz}{$f_{\rm esc}(z)$}
\newcommand{\ghi}{$\Gamma_{\rm H \,{\scriptscriptstyle I}}$}
\newcommand{\hhi}{$\xi_{\rm H \,{\scriptscriptstyle I}}$}
\newcommand{\ghiz}{$\Gamma_{\rm H\,{\scriptscriptstyle I}}(z)$}
\newcommand{\ghei}{$\Gamma_{\rm He \,{\scriptscriptstyle I}}$}
\newcommand{\hhei}{$\xi_{\rm He \,{\scriptscriptstyle I}}$}
\newcommand{\gheiz}{$\Gamma_{\rm He \,{\scriptscriptstyle I}}(z)$}
\newcommand{\gheii}{$\Gamma_{\rm He \,{\scriptscriptstyle II}}$}
\newcommand{\hheii}{$\xi_{\rm He \,{\scriptscriptstyle II}}$}
\newcommand{\gheiiz}{$\Gamma_{\rm He \,{\scriptscriptstyle II}}(z)$}
\newcommand{\afuv}{$A_{\rm FUV}$}
\newcommand{\afuvz}{$A_{\rm FUV} (z)$}

\begin{abstract} 
We present new synthesis models of the extragalactic background light (EBL) 
from far infra-red (FIR) to TeV $\gamma$-rays, with an emphasis on the extreme 
ultraviolet (UV) background which is responsible for the observed ionization 
and thermal state of the intergalactic medium across the cosmic time. Our 
models use updated values of the star formation rate density and dust attenuation 
in galaxies, QSO emissivity, and the distribution of \hi~gas 
in the IGM. Two of the most uncertain parameters in these models, the escape 
fraction of \hi~ionizing photons from galaxies and the spectral energy distribution 
(SED) of QSOs, are determined to be consistent with the latest measurements of 
\hi~and \heii~photoionization rates, the \heii~Lyman-$\alpha$ effective optical depths, 
various constraints on \hi~and \heii~reionization history and many measurements 
of the local EBL from soft X-rays till $\gamma$-rays. We calculate the EBL from FIR to 
TeV $\gamma$-rays by using FIR emissivities from our previous work and constructing 
an average SED of high-energy emitting QSOs, i.e, type-2 QSOs and blazars. For public 
use, we also provide the EBL models obtained using different QSO SEDs at extreme-UV 
energies over a wide range of redshifts. These can be used to quantify uncertainties 
in the parameters derived from photoionization models and numerical simulations 
originating from the allowed variations in the UV background radiation.
\end{abstract}


\section{Introduction}\label{sec1}
The radiation background set-up by light emitted from all galaxies and Quasi-Stellar 
Objects (QSOs) throughout the cosmic time is known as extragalactic background light 
(EBL). The full spectrum of EBL carries imprints of cosmic structure formation, 
therefore, it serves as an important tool to study formation and evolution of galaxies.
It is also essential for studying the propagation of high energy $\gamma$-rays from 
distant sources since the EBL can annihilate $\gamma$-rays upon collision 
\citep{Gould66, Stecker92, GammaSci}. The EBL is playing a key role in rapidly 
developing $\gamma$-ray astronomy to address fundamental questions related to the 
production of $\gamma$-rays, their acceleration mechanism \citep{Stecker07} and cosmic 
magnetic fields in the intergalactic space 
\citep{Neronov10, Tavecchio11, Arlen12, Finke15}.  

A small part of this EBL at extreme-UV energies (${\rm E}>13.6$ eV or $\lambda<912$ \AA) 
is known as the UV background (UVB). The UVB is responsible for maintaining the observed 
ionization and thermal state of the diffuse intergalactic medium 
\citep[IGM; see reviews by][]{Meiksin09, McQuinn16}, a reservoir that is believed to 
contain more than 90\% of total baryons in the Universe.

The sources that setup UVB also drive the major phase transitions of the IGM, the \hi~and 
\heii~reionization. Starting from the time when first galaxies were born, the process of 
\hi~reionization is believed to be completed around $z\sim6$ as suggested by various 
observations of \hi~Lyman-$\alpha$ forest in the spectra of high redshift QSOs 
\citep[e.g.,][]{Beckerr01, Fan06, Goto11, Mcgreer15, Greig17, Banados17, Davies18}, electron 
scattering optical depth to the cosmic microwave background \citep[CMB][]{Larson11, Planck16} 
and decreasing fraction of Lyman-$\alpha$ emitting galaxies at high redshifts 
\citep[e.g.,][]{Schenker14, Choudhury15, Mesinger15, Mason17}. The \heii~reionization is 
believed to be started by hard \heii~ionizing radiation emitted by QSOs and took
longer time to complete. It was completed around $z\sim2.8$ as suggested by observations 
of \heii~Lyman-$\alpha$ forest in handful of QSOs 
\citep[e.g.,][]{Kriss01, Shull04, Shull10, Fechner06, Worseck11, Worseck16} and 
measurements of a peak in the redshift evolution of IGM temperature 
\citep{Lidz10, Becker11t, Hiss17, Walther18}. During these reionization events, spectrum of the UVB 
decides the extra energy gained by photoelectrons, which gets redistributed in the IGM 
driving its thermal state \citep{Hui97}. Subsequently the thermal history of the IGM is 
driven by the UVB and adiabatic expansion of the Universe.

Thermal history of the IGM has its imprint on the small scale structures in the IGM 
through pressure smoothing that can be probed by correlation analysis of closely spaced QSO 
sightlines \citep[e.g.,][]{Gnedin98, Schaye01, Kulkarni15, Rorai17}. Photoheating by the 
UVB can also provide negative feedback that can suppress the star formation in dwarf galaxies 
\citep{Efstathiou92, Weinberg97} and decide the faint end shapes of the high-z luminosity 
functions \citep[e.g.,][]{Samui07}. Therefore the UVB is one of the most important inputs 
in the cosmological simulations of structure formation and the IGM 
\citep[e.g,][]{Hernquist96, Dave99, Springel01}. 
 
Perhaps the most important and frequent application of UVB is to study the metal absorption 
lines ubiquitously observed in QSO absorption spectra. It is because the EBL from 
extreme-UV to soft X-ray not only ionizes hydrogen and helium but also several metals such as 
Mg, C, Si, N, O, Ne. Therefore, the UVB serves as an essential ingredient to study the 
physical and chemical properties of the gas observed in the QSO absorption spectra 
originating either from low-density gas in the IGM or high-density gas in the vicinity of 
intervening galaxies known as a circumgalactic medium (CGM). In the studies of metal 
absorption lines, the UVB is crucial for relating observed ionic abundances to metal 
abundances, in order to determine the metal production and their transport to CGM by 
galaxies \citep[e.g.,][]{Ferrara05, Lehner14, Peeples14} and the time evolution of the 
cosmic metal density \citep[e.g.,][]{Songaila96, Bergeron02, Schaye03, Aracil04, 
Dodorico13, Shull14, Prochaska17, Muzahid17}. At low redshifts, the UVB is essential for 
studying the ionization mechanism of highly ionized species such as O~{\sc vi} 
\citep[e.g.,][]{Danforth05, Tripp08, Muzahid12o6, Savage14, Pachat16, Narayanan18} and 
Ne~{\sc viii} \citep[e.g.,][]{Savage05, Savage11, Narayanan12, Meiring13, Hussain15, 
Hussain17, Pachat17} to understand their contribution to the warm-hot IGM and missing 
baryons \citep[see][]{Shull12}.

The local EBL ($z=0$) at most wavelength ranges can be observed directly 
\citep[e.g.,][]{Dwek98, Dole06, Ajello08}, however there are no such direct observations 
of the UVB because the interstellar-medium of our Milky-way attenuates it completely. 
Therefore, one needs to model the UVB spectrum and its redshift evolution. There are, 
however, integral constraints on the UVB obtained from the measurements of \hi~and 
\heii~photoionization rates. The \hi~photoionization rates (\ghi) can be measured by using 
the observations of 21 cm truncation and H$\alpha$ fluorescence in nearby galaxies 
\citep[e.g.,][]{Sunyaev69, Dove94, Adams11, Fumagalli17}, by analyzing the incidence of
Lyman-$\alpha$ forest lines in the proximity of QSOs \citep[e.g.,][]{Bajtlik88, Kulkarni93, Srianand96, Dallaglio08}, 
and by reproducing various statistical properties of the observed Lyman-$\alpha$ forest 
in the cosmological simulations of the IGM where \ghi~is treated as one of the free 
parameters \citep[e.g.,][]{Rauch97, Bolton07, Becker13, Kollmeier14, Shull15, Gaikwad17a, Gaikwad17b, Gaikwad17c}. 
One can also infer the \heii~photoionization rate (\gheii) using the measurements of 
\ghi~and the observed \hi~column density distribution as demonstrated in \citet{Khaire17sed}. 
These measurements play crucial role in calibrating the synthesis models of UVB at different 
redshifts. 

The full synthesis model of UVB was pioneered  by \citet{HM96} 
\citep[see also][]{Fardal98} using cosmological radiative transfer calculations following 
the footsteps of previous work \citep{Miralda90, Shapiro94, Giroux96}. Over the last two 
decades, there are some variations of their UVB models \citep{HM01, HM12} and the UVB models 
by other groups \citep{Shull99, FG09}. Out of these, most recent models\footnote{Excluding a 
QSO only model by \citet[][]{Madau15} and a recent \citet{Puchwein18} model which appeared 
while we were finalizing the paper.} are \citet[][hereafter \citetalias{FG09}]{FG09} and 
\citet[][hereafter \citetalias{HM12}]{HM12}. These recent models are not completely consistent 
with the new observations such as the \ghi~at $z<0.5$ \citep{Shull15, Gaikwad17a, Gaikwad17b, Viel17, Gurvich17} 
and $z>3$ \citep{Becker13}. These models also use old values of many observables relevant to 
UVB which are significantly different from current measurements, such as, the type-1 QSO emissivity 
\citep[][hereafter \citetalias{Khaire15puc}]{Khaire15puc}, galaxy emissivity 
\citep[][hereafter \citetalias{Khaire15ebl}]{Behroozi13, Madau14, Khaire15ebl} and various 
constraints on \hi~and \heii~reionization \citep[][]{Choudhury15, Planck16b, Greig17b, Worseck16}. 
In light of all these issues, and the fact that there are only few independent UVB models available in the 
literature, we present new synthesis models of EBL focusing on the UVB and extending it 
till TeV $\gamma$-rays from far infra-red (FIR).

In our EBL models we use the updated type-1 QSO emissivity obtained from a compilation of 
recent QSO luminosity functions \citetalias{Khaire15puc}. Our models use galaxy emissivity 
till FIR wavelengths obtained from the updated star formation and dust attenuation 
history of the Universe from \citetalias{Khaire15ebl}. These are determined through a 
large compilation of multi-wavelength galaxy luminosity functions 
\citepalias[see][and references therein]{Khaire15ebl}. We also use updated \hi~distribution 
of the IGM from \citet{InoueAK14} that is obtained from a large number of different
observations (mentioned in Section~\ref{sec2}). Apart from these, there are two additional 
important input parameters required to model the UVB; the average escape fraction of 
\hi~ionizing photons from galaxies ($f_{\rm esc}$) and the mean spectral energy distribution 
(SED) of type-1 QSOs at extreme-UV wavelengths. However, the observational constraints on 
these are either not available or poor. 

The measurement of $f_{\rm esc}$ from individual galaxies as well as large surveys has been 
proven to be a challenging endeavor \citep{Vanzella10, Siana15, Mostardi15}. There are 
handful of galaxies at $z<0.5$ which show emission of \hi~ionizing photons with $f_{\rm esc}$ 
ranging from 2 to 46\% \citep{Bergvall06, Leitet13, Borthakur14, Leitherer16, Puschnig17, Izotov16a, Izotov16b, Izotov18} 
and only two galaxies at $z>3$ with large $f_{\rm esc}$ of the order of 50\% 
\citep{Vanzella16, deBarros16, Shapley16}. However, all of these are few exceptional cases, 
as most studies with large number of galaxies provide only upper limits on the average $f_{\rm esc}$ 
\citep[e.g.,][see left-hand panel of Fig.~\ref{fig3} and Table~\ref{tab.escape} for summary of 
recent measurements]{Cowie09, Bridge10, Siana10, Guaita16, Matthee17, Grazian17, Japelj17, Smith18}.
On the other hand, there are measurements of QSO SED at extreme-UV from large number of QSOs 
probing rest-wavelength $\lambda <912$ 
\AA~\citep[e.g,][]{Zheng97, Telfer02, Scott04, Shull12, Stevans14, Lusso15, Tilton16}. However, 
under the assumption that QSO SED follows a power-law, 
$f_{\nu} \propto \nu^{\alpha}$ at $\lambda< 912$~\AA~, the values of power-law index $\alpha$ 
obtained from these measurements show large variation 
\citep[-0.56 to -1.96; see table 1 of][for the summary of these measurements]{Khaire17sed}. 
Also, the smallest wavelength probed in these studies is $\sim 425$~\AA~whereas the UVB model 
calculations extrapolate it upto soft X-rays ($\lambda \lesssim 20$~\AA). 

Because of these uncertainties, we choose to determine the $f_{\rm esc}(z)$ and $\alpha$ which 
are required to consistently reproduce various observational constraints on the UVB following 
\citet{Khaire16} and \citet{Khaire17sed}. We obtain the $f_{\rm esc}(z)$ to reproduce the 
\ghi~measurements and various constraints on \hi~reionization. In our fiducial UVB model, we 
use $\alpha=-1.8$ which was found to reproduce the measured \heii~Lyman-$\alpha$ effective 
optical depths as a function of $z$ and the epoch of \heii~reionization \citep[see][]{Khaire17sed}. 
However, we also provide UVB models for $\alpha$ varying from -1.4 to -2.0 in the interval of 
0.1. Moreover, following \citet{Sazonov04} we modify the SED of type-1 QSOs to include the 
type-2 QSOs and blazars in order to calculate the X-ray and $\gamma$-ray part of the EBL 
consistent with various measurements of the local X-ray and $\gamma$-ray backgrounds. Our full 
EBL spans more than fifteen orders of magnitude in wavelength from FIR to TeV $\gamma$-rays. 

The paper is organized as follows. In Section~\ref{sec2}, we briefly discuss the basic 
cosmological radiative transport theory used for calculating the EBL. In Section~\ref{sec3}, 
we explain the emissivities used in our EBL. We discuss the QSO emissivity, their SEDs, 
the galaxy emissivity, \fescz~ and the diffuse emissivity from the IGM. In Section~\ref{sec4}, 
we discuss our fiducial model and its predictions for  \hi~and \heii~photoionization rates,
reionization histories, full spectrum of the EBL from FIR to $\gamma$-rays and the detailed 
uncertainties in the UVB models. In Section~\ref{sec5}, we summarize our main results. 
In Appendix we show plots for various UVB models generated for different $\alpha$, provide 
relevant tables of photoionization and photoheating rates and optical depths encountered by 
$\gamma$-ray photons due to the EBL. Throughout the paper, we have used the cosmological 
parameters $\Omega_{m}=0.3$, $\Omega_{\Lambda}=0.7$ and $H_0=70$ km s$^{-1}$ Mpc$^{-1}$ 
consistent with measurements from \citet{Planck16}. All our EBL tables in machine-readable format 
are publicly available to download (at \href {http://www.iucaa.in/projects/cobra/} {cobra-webpage} \& a
\href {http://vikramkhaire.weebly.com/downloads.html} {homepage})
and latest version of {\sc cloudy} software \citep[last described in][]{Ferland17}.

\section{Cosmological Radiative transfer}\label{sec2}
The specific intensity, $J_{\nu_0}$,  of the EBL (in units of erg cm$^{-2}$ s$^{-1}$ Hz$^{-1}$ 
sr$^{-1}$) at frequency $\nu_0$ and redshift $z_0$ is obtained using the following integral 
\citep[]{Peebles93, HM96}:
\begin{equation}\label{rad_t}
J_{\nu_{0}}(z_{0})=\frac{c}{4\pi}\int_{z_{0}}^{\infty}dz\,\frac{(1+z_{0})^{3}
\,\epsilon_{\nu}(z)}{(1+z)\, {\rm H}(z)} \, {\rm e}^{-\tau_{\rm eff}(\nu_{0},\, z_{0},\, z)}.
\end{equation}
Here, $c$ is speed of light, $ {\rm H} (z) = H_0 \sqrt{\Omega_m(1+z)^3+\Omega_{\Lambda}}$ is the 
Hubble parameter, $\epsilon_{\nu} (z)$ is a volume averaged comoving specific emissivity at 
frequency $\nu$ with $\nu=\nu_{0}(1+z)/(1+z_{0})$ and $\tau_{\rm eff}(\nu_{0}, z_{0}, z)$ 
is an effective optical depth encountered by photons arriving at redshift $z_0$ having frequency 
$\nu_0$ which were emitted at redshift $z \ge z_0$ with frequency $\nu$. The $\tau_{\rm eff}$ from 
the IGM is negligibly small for EBL at optical and higher wavelengths. However, it can be 
significantly larger for extreme-UV radiation due to atomic gas in the IGM. Therefore, it needs to 
be accurately calculated using detailed radiative transfer through hydrogen and helium gas of the 
IGM present in the form of discrete absorbers. 

The effective optical depth is defined as $\tau_{\rm eff}=-{\rm ln} (<e^{-\tau}>)$ where $\tau$ 
is continuum optical depth through absorbers and $<e^{-\tau}>$ is an average transmission over all 
lines of sights.  Assuming that the discrete absorbers along any line of sight are Poisson 
distributed, the $\tau_{\rm eff}$ is obtained by \citep[see][]{Paresce},
%
\begin{equation}\label{taueff}
\tau_{\rm eff}(\nu_{0}, z_{0}, z)=\int_{z_{0}}^{z}dz'\int_{0}^{\infty}
dN_{\rm HI}\frac{\partial^{2}N}{\partial N_{\rm HI}\,\partial z'}(1-{\rm e}^{-\tau_{\nu'}}).
\end{equation}
%
Here, $\partial^{2}N/\partial N_{\rm HI}\partial z'= f(N_{\rm HI}, z')$ is a bivariate distribution of 
the absorbers with respect to their redshift $z'$ and \hi~column density $N_{\rm HI}$, also known as 
a column density distribution function of \hi. The $\tau_{\nu '}$ in equation~(\ref{taueff}) is a
continuum optical depth encountered by photons arriving at redshift $z_0$ with frequency $\nu_0$ 
which were emitted at redshift $z'\ge z_0$ with frequency $\nu'=\nu_{0}(1+z')/(1+z_{0})$. 
The $\tau_{\nu '}$, by ignoring a negligible contribution from metals and dust in the IGM, is 
given by 
%
\begin{equation}\label{tauc}
\tau_{\nu'}=N_{\rm HI}{\sigma_{\rm HI}(\nu')}+N_{\rm HeI}{\sigma_{\rm HeI}(\nu')} 
+N_{\rm HeII}{\sigma_{\rm HeII}(\nu')},
\end{equation} 
%
where $N_x$ and $\sigma_x$ are the column densities and photoionization cross-sections of species 
$x$. Unlike for \hi, the column density distribution of \heii~and \hei~are not 
available from direct observations.

There are very few lines of sight where \nheii~has been measured \citep[e.g.,][]{Zheng04, Muzahid11, McQuinn14}. 
Therefore, the amount of \nhei~and \nheii~in the \hi~absorbers needs to be inferred
through photoionization modeling. For that we rewrite the equation~(\ref{tauc}) as,
%
\begin{equation}\label{tauc_mod}
\tau_{\nu'}=N_{\rm HI}\Big[{\sigma_{\rm HI}(\nu')}+\zeta{\sigma_{\rm HeI}(\nu')} 
+\eta{\sigma_{\rm HeII}(\nu')}\Big],
\end{equation}
%
where $\zeta=N_{\rm HeI}/N_{\rm HI}$ and $\eta=N_{\rm HeII}/N_{\rm HI}$. Under the assumption that 
the IGM absorbers are in photoionization equilibrium with the UVB, $\eta$ is obtained by solving 
following quadratic equation \citep{Fardal98, FG09, HM12}
%
\begin{equation}\label{quad}
\begin{aligned}
\frac{n_{\rm He}}{4n_{\rm H}}\,
\frac{\Gamma_{\rm HI}}{n_e\alpha_{\rm HI}(\rm T)}\,
\frac{\sigma_{912}N_{\rm HI}}{(1+{\rm A}\sigma_{912}N_{\rm HI})}
=\sigma_{228}N_{\rm HeII}\\
\,+\,\frac{\Gamma_{\rm HeII}}{n_e \alpha_{\rm HeII}(\rm T)}\,
\frac{\sigma_{\rm HeII}N_{\rm HeII}}{(1+{\rm B}\sigma_{228} N_{\rm HeII})}.
\end{aligned}
\end{equation}
%
Here, $n_e$ is the electron density, $\sigma_{228}$ and $\sigma_{912}$ are the photoionization 
cross-sections of \heii~and \hi~at 228 \AA~and 912 \AA, respectively, the constants A and B are obtained 
to fit the numerical results and $\Gamma_x$ is the photoionization rate for species $x$. The $\Gamma_x$ 
at any redshift $z_0$ is defined as
%
\begin{equation}\label{eq.gama}
\Gamma_{x}(z_0)=\int_{\nu_{x}}^{\infty}d\nu\,
\frac{4\pi\,J_{\nu}(z_0)}{h\nu}\,\sigma_{x}(\nu)\,\, ,
\end{equation}
%
where, $\nu_x$ is a threshold frequency for ionization of species $x$. The ionization threshold energies 
$h\nu_{x}$, where $h$ is the Planck's constant, for \hi, \hei~and \heii~are 13.6, 24.4 and 54.4 eV, 
respectively. Following \citet{HM12} we take ${\rm A=0.02}$ and ${\rm B=0.25}$, the relation between $n_e$ and 
$N_{\rm HI}$ as $n_e=1.024\times10^{-6} (N_{\rm HI}\Gamma_{\rm HI})^{(2/3)}{\rm cm^{-3}}$ and T=20000 K. 
We use $\zeta= \eta n_e \, \alpha_{\rm He I}{\rm (T)}/{\Gamma_{\rm He I}}$. We verified these parameters 
by modeling the IGM clouds as plane parallel slabs with the line-of-sight thickness equal to the Jeans 
length \citep{Schaye01} using {\sc cloudy13} software \citep{Ferland13}. 

We take $f(N_{\rm HI}, z)$ from \citet{InoueAK14} at all $z$. It has been obtained by fitting various 
observations over $z=0-6$ such as the number distribution and column density distribution of 
optically thin \hi~ Lyman-$\alpha$ absorbers \citep{Weynmann98, Kim01, Janknecht06, Kim13}, optically 
thick Lyman limit absorbers \citep{Peroux05, Rao06, Songaila10, Prochaska10, OMeara13, Fumagalli13} 
and damped Lyman-$\alpha$ absorbers \citep{OMeara07, Noterdaeme09, Noterdaeme12, Prochaska14}, the 
mean transmission and optical depth of the IGM to Lyman-$\alpha$ photons 
\citep{FG08, Kirkman07, Fan06, Becker13Ly} and the mean free path of \hi~ionizing photons through 
IGM \citep{Prochaska09, Worseck14LLS}. The $f(N_{\rm HI}, z)$ calculated at $z=5.5$ and 6 from 
high-resolution hydrodynamic simulations for the measured \ghi~values match reasonably well with the  
$f(N_{\rm HI}, z)$ from \citet{InoueAK14}, as shown in \citet{Khaire16}. This consistency motivates 
us to use this $f(N_{\rm HI}, z)$ even at $z>6$ where direct observations are not possible because of 
strong Gunn-Peterson effect \citep{Gunn65}. See Section~\ref{sec4.4.3}, for uncertainties in the UVB 
arising when different $f(N_{\rm HI}, z)$ is used in our calculations. 

\section{Emissivity}\label{sec3}
The comoving source emissivity $\epsilon_{\nu}$ is the most important quantity in the EBL calculations. 
The $\epsilon_{\nu}$ is contributed by photons emitted from three different sources; the QSOs 
($\epsilon^Q_{\nu}$) by accretion of matter onto supermassive blackholes, the galaxies ($\epsilon^G_{\nu}$) 
from the radiation emitted by their stellar population and re-processed by dust in their interstellar medium, 
and the re-processed diffuse emission from the photoionized IGM ($\epsilon^d_{\nu}$). Therefore,
%
\begin{equation}\label{zeta}
\epsilon_{\nu}(z)=\epsilon^Q_{\nu}(z)+\epsilon^G_{\nu}(z)+\epsilon^d_{\nu}(z).
\end{equation}
%
In the following subsections, we discuss the emissivity from these different sources used in our EBL 
calculations. Note that, we do not consider non-standard sources such as dark-matter annihilation which has 
been shown to have negligible contribution to the EBL \citep[e.g.,][\citetalias{Khaire15puc}]{Zavala11, Ajello15, Gaikwad17a}.
%
%
\begin{figure*}
\centering
\includegraphics[totalheight=0.745\textheight, trim=0.0cm 0.2cm 7.5cm 0.0cm, clip=true, angle=270]{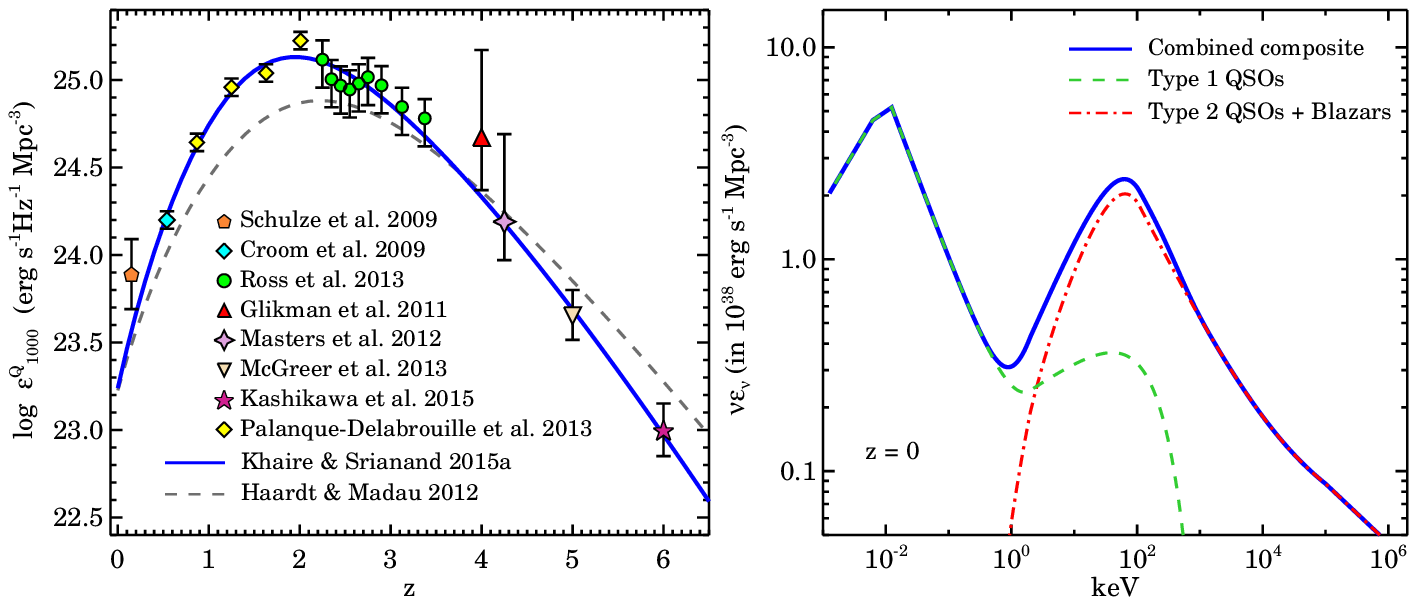}
\caption{
Left-hand panel: the specific QSO emissivity at $1000$ \AA~($\epsilon^Q_{\nu_{1000}}$) 
with $z$. Data points are taken from the compilation of recent QSO luminosity functions
by \citet[][see their table 1]{Khaire15puc}. Blue solid 
curve is a simple fit through the $\epsilon^Q_{\nu_{1000}}$ values (equation~\ref{eqso}). 
Dashed curve is the $\epsilon^Q_{\nu_{1000}}$ used by \citetalias{HM12} which is 
obtained using the compilation of QSO luminosity functions by \citet{Hopkins07}. 
Right-hand panel: $z=0$ QSO emissivity (solid curve) at different energies to
illustrate our fiducial composite QSO SED (with $\alpha=-1.8$; equation~\ref{sed}). 
Dashed curve and dot-dash curve shows the adopted SED of Type-1 QSOs and high-energy 
emitting AGN template SED (i.e, Type-2 QSOs and blazars), respectively, normalized to get 
$\epsilon^Q_{\nu_{1000}}$ at $z=0$. 
}
\label{fig1}
\end{figure*}
%
%
\subsection{QSO Emissivity}\label{sec3.1}
The QSO emissivity $\epsilon^Q_{\nu_0}(z)$ is obtained by integrating the QSO luminosity function (QLF), 
$\phi(L_{\nu_0}, z)$, observed at frequency $\nu_0$ and redshift $z$ by
%
\begin{equation}\label{eq.rho}
\epsilon^Q_{\nu_0}(z)=\int_{L_{\nu_0}^{\rm min}}^{\infty} {L_{\nu_0}(z) \phi(L_{\nu}, z) dL_{\nu_0}}\,,
\end{equation}
%
where $L_{\nu_0}^{\rm min}$ is the minimum luminosity of QSOs at $\nu_0$ used in the integration. 
This $\epsilon^Q_{\nu_0}(z)$ can be used to obtain the QSO emissivity $\epsilon^Q_{\nu}(z)$ at any 
frequency $\nu$ by using the mean SED of QSOs ($k_{\nu}$) as 
%
\begin{equation}\label{eqso}
\epsilon^Q_{\nu}(z)=k_{\nu}*\epsilon^Q_{\nu_0}(z).
\end{equation}
%
We use $\epsilon^Q_{\nu_0}(z)=\epsilon^Q_{1000}(z)$ obtained at $\nu_0=c/1000$ \AA~(or $h\nu_0=12.4$ eV) from 
\citetalias{Khaire15puc}. In \citetalias{Khaire15puc}, the $\epsilon^Q_{\nu_0}(z)$ values were obtained 
by using a compilation of recent QLFs (of type-1 QSOs) at different wavebands 
\citep{Schulze09, Croom09, Glikman11, Masters12, Ross13, Palanque13, McGreer13, Kashikawa15}. 
For each QLF, $\epsilon^Q_{\nu_0}$ was obtained at the observed frequency $\nu_0 < c/1000 $ \AA, by using 
equation~(\ref{eq.rho}) with $L_{\nu}^{\rm min}=0.01L_{\nu_0}^*$, where $L_{\nu_0}^*$ is the 
characteristic luminosity of that QLF. Then these $\epsilon^Q_{\nu_0}$ values 
\citepalias[see Table 1 of][]{Khaire15puc} were converted at frequency $c/1000$ \AA~using a form of QSO 
SED applicable at $\lambda \ge 1000$\AA. A simple fit through the redshift evolution of 
$\epsilon^Q_{1000}(z)$ in units of ${\rm erg\, s^{-1}\, Hz^{-1} Mpc^{-3}}$ is given by\footnote{The 
$\epsilon^Q_{1000}(z)$ is obtained by multiplying $(1000/912)^{1.4}$ to the $\epsilon^Q_{912}(z)$ provided 
in \citetalias{Khaire15puc}},
%
\begin{equation}\label{eqso}
\epsilon^Q_{1000}(z)=4.53\times 10^{24} \, (1+z)^{5.9}\,\frac{\exp(-0.36z)}{\exp(2.2z) + 25.1}\,\,.
\end{equation}
%
%
In the left-hand panel of Fig.~\ref{fig1}, we show this fit with the individual $\epsilon^Q_{1000}(z)$ 
values for each QLF compiled in \citetalias{Khaire15puc}. For comparison, we also show the 
$\epsilon^Q_{1000}(z)$ used by \citetalias{HM12} which was obtained using the QLFs 
compiled by \citet{Hopkins07}. At $z<3$ the $\epsilon^Q_{1000}(z)$ by \citetalias{Khaire15puc} is higher 
than \citetalias{HM12} because the recent QLFs \citep[e.g.,][]{Croom09, Palanque13} have higher 
$L_{\nu_0}^*$ and characteristic number density $\Phi_{\nu_0}^*$. This higher QSO emissivity was crucial in 
resolving the photon underproduction crisis claimed by \citet{Kollmeier14}. We are using the same SED 
at $\lambda>1000$\AA~as used in \citetalias{Khaire15puc}, however, at $\lambda<1000$\AA~we use a modified 
form of the SED. 

The QSO emissivity mentioned in equation~(\ref{eqso}) is contributed by type-1 QSOs alone. Note that 
to calculate the background radiation in extreme-UV, one does not need to consider the contribution 
from type-2 QSOs. According to the standard unification scheme of active galactic nuclei 
\citep[AGN;][]{Antonucci93, Urry95}, different classes of QSOs arise due to differences in the orientation of
QSOs along the direction of obscuring torus around their central engine with respect to us. Therefore, 
under the assumption of isotropic distribution of randomly oriented QSOs, to calculate the extreme-UV 
emissivities it is equivalent to assume that the extreme-UV photons are emitted either isotropically by 
type-1 QSOs alone or only along certain directions by both type-1 and type-2 QSOs. However, in the 
latter assumption one needs to correctly account for the fraction of type-2 QSOs. For simplicity, to 
calculate the UVB we choose the former assumption. However, to extend our EBL calculation to high energy 
X-rays, which are emitted isotropically by all types of QSOs, we need to incorporate contribution from 
type-2 QSOs.

This contribution in X-rays can be accounted self-consistently by using type-2 QSO luminosity function in 
soft and hard X-ray band and then modeling the distribution of different hydrogen column densities in 
obscuring torus, using intrinsic template of X-ray SEDs of QSOs and by fixing a contribution of extremely 
obscured Compton thick QSOs by comparing with measurements of the unresolved X-ray background 
\citep[see for e.g.,][]{Comastri95,Treister05,Gilli07,Treister09,Ballantyne11}. Similarly the contribution 
to $\gamma$-ray background from blazars can be modeled by generating the blazar luminosity function and 
their SED from luminosity functions of radio QSOs \citep[e.g.,][]{Draper09}.

However, these X-ray population synthesis models require different fractions of Compton thick QSOs to 
generate the observed peak at $\sim 30$ keV in X-ray background and varying degree of obscuring hydrogen 
column densities but still may not entirely reproduce the observed soft X-ray background 
\citep[see,][]{Cappelluti17}. Although this might be the most appropriate approach, given the large 
uncertainties involved in it we take a more simplistic approach which can provide an observationally 
consistent model of the EBL in X-ray energies useful for constructing ionization models for highly 
ionized metal species such as O~{\sc vi},  O~{\sc vii}, Ne~{\sc viii}, Ne~{\sc ix}, Mg~{\sc x}. In 
particular, we follow the approach of \citet{Sazonov04} and construct a template SED of type-1 
and type-2 QSOs at X-ray energies. Our type-2 QSO SED not only accounts for the contribution of Compton thick 
AGNs but also includes $\gamma$-rays that will mostly come from beamed sources like blazars. This SED can 
be thought as a template SED for high-energy emitting AGNs. These SEDs are constructed such that when 
used in the EBL calculations they can reproduce a complete spectrum of observed X-ray and $\gamma$-ray 
background. The final QSO SED constructed in this way is given below.

Our final composite QSO SED ($k_{\nu}$) is a combination of the SEDs from type-1 
($k^{\rm Q1}_{\nu}$) and type-2 QSOs including blazars ($k^{\rm Q2}_{\nu}$). Therefore, 
\begin{equation}\label{sed}
 k_{\nu}=k^{\rm Q1}_{\nu} + k^{\rm Q2}_{\nu}
\end{equation}
We use a following SED for the type-1 QSOs:  
\begin{equation}\label{sed1}
k^{\rm Q1}_{\nu}=
\begin{cases}
A_1 \Big( \frac{h\nu}{6.2 {\,\,\rm eV}} \Big)^{-0.5}, & h\nu < {\rm 6.2 \,\,eV} \\
A_2 \Big( \frac{h\nu}{12.4 {\,\,\rm eV}} \Big)^{-0.8}, & 6.2 < h\nu < {\rm 12.4 \,\,eV} \\
A_3 \Big( \frac{h\nu}{12.4 {\,\,\rm eV}} \Big)^{\alpha} \exp \Big( \frac{\nu}{\nu_a} \Big), & 12.4 {\rm \,\,eV}< h\nu < h\nu_a  \\
A_4 \Big( \frac{\nu}{\nu_a} \Big)^{-0.8} \exp \Big(\frac{-h\nu}{{\rm 2 \,\,MeV}} \Big), & h\nu > h\nu_a 
\end{cases}
\end{equation}
where the normalizations are 
\[
\begin{aligned}
 A_1=2^{0.8},  \,\,\,  A_2=1, \,\,\,  A_3=\exp\Big(\frac{-12.4{\,\,\rm eV}}{h\nu_a}\Big), \\
 A_4=A_3 \exp\Big(1 + \frac{h\nu_a}{2 {\rm \,\,MeV}}\Big)\Big( \frac{h\nu_a}{12.4 {\,\,\rm eV}} \Big)^{\alpha}.
\end{aligned}
\] 
The power-law SED form and the indices at $h\nu< 12.4$ eV ($\lambda > 1000$ \AA) are taken from the 
measurements of \citet{Stevans14}. At ionizing energies, $h\nu> 12.4$ eV, we consider a range in 
power-law index $\alpha$ for our UVB models since there is no consensus on the observed value of 
$\alpha$, as mentioned in Section~\ref{sec1}. This power-law SED has been multiplied by $\exp({\nu/\nu_a})$ 
to get the excess in soft X-rays and the hard X-ray bump. However, the value of $\nu_a$ depends on the 
assumed value of $\alpha$. Therefore for each $\alpha$ we have used a different $h\nu_a$ as given in 
Table~\ref{t1}. We calculated EBL by varying $\alpha$ in the interval of 0.1 from $-1.4$, consistent with 
low-z ($z<1.5$) measurements of \citet{Stevans14}, to $\alpha=-2$, consistent with high-z ($z\sim 2.5$) 
measurements of \citet{Lusso15} and from radio-loud QSO sample of \citet{Telfer02}. For our fiducial UVB 
model, we use $\alpha=-1.8$ that has been shown to be consistent with the \heii~Lyman-$\alpha$ effective 
optical depth measurements \citep{Khaire17sed}. In the right-hand panel of Fig.~\ref{fig1}, we show the 
type-1 QSO SED for $\alpha=-1.8$ normalized at $\epsilon^Q_{1000}$ at $z=0$. 

For type-2 QSOs and $\gamma$-ray emitting blazars, we adopt the following SED, 
\begin{equation}\label{sed2}
k^{\rm Q2}_{\nu}=
\begin{cases}
B_1 \, p_{\nu} & h\nu < E_0 \\
B_2 \,  p_{\nu} \Big( \frac{h\nu}{{E_0}} \Big)^{-0.24} \exp \Big(\frac{-h\nu}{{\rm 83 \,\,keV}} \Big), & E_0 < h\nu < E_1  \\
B_3 \,  p_{\nu} \Big( \frac{h\nu}{E_1} \Big)^{-1.6}  \, \Big[1+q\,(\frac{h\nu}{\rm 1 \,\,keV})^{0.54}  \Big], & E_1 < h\nu <  E_2\\
B_4 \,  p_{\nu} \Big( \frac{h\nu}{E_2} \Big)^{-1.28} \exp \Big(\frac{-h\nu}{{\rm 600 \,\,GeV}} \Big), & h\nu > E_2 
\end{cases}
\end{equation}
where
\[
 p_{\nu} =S_k \exp \Big(-\frac{{\rm 1 \,\,keV}}{h\nu} \Big),  \,\,\,  q=4.1\times 10^{-3},   \,\,\, 
\]
\[
 E_0={\rm 2 \,\,keV}, \,\,\, E_1={\rm 113 \,\,keV} \,\,\, {\rm and \,\,\,} E_2={\rm 100\,\,MeV}.
\]
Here the factor $S_k$ has been adjusted to match the hard X-ray background measurements at $z=0$. Its value 
depends on the value of assumed $\alpha$. The $S_k$ values for different $\alpha$ are also given in 
Table~\ref{t1}. The normalizations in equation~(\ref{sed2}) are 
\[
 B_1=A_3 \Big(\frac{E_0}{12.4 {\,\, \rm eV}} \Big)^{\alpha} \exp \Big( \frac{E_0}{h\nu_a} \Big),  \,\,\,  
 B_2=B_1 \exp\Big(\frac{2}{83} \Big), \,\,\, 
\] 
\[ 
 B_3=9.246 \times 10^{-2} B_2 \,\,\, {\rm and} \,\,\, 
 B_4=5.444 \times 10^{-6} B_2.
\] 
Note that with these normalization factors (see $B_1$) along with the $S_k$, the type-2 QSO SED has been 
scaled with type-1 QSO emissivity. The form of type-2 QSO SED has been adopted from  \citet{Sazonov04} which 
we modified with different normalizations and a very high-energy part that reproduces the $z=0$ X-ray and 
$\gamma$-ray background measurements up to TeV energies (see Section~\ref{sec4.3} and Fig.~\ref{figA3}). Our 
type-2 QSO SED even includes $\gamma$-ray emission, therefore it serves as a combination of all high-energy 
emitting type of QSOs.
%
%
\begin{table}
\caption{Parameters of QSO SED given by equation~\ref{sed1} and \ref{sed2}}
\def\arraystretch{1.5}
\begin{tabular}{ l c c c  }              
\hline
 Model Name   & $\alpha$     & $h \nu_a$ (keV)   &  $S_k$  \\
\hline      
Q14     &      $ -1.4 $      & 40       & 0.4 \\
Q15     &      $ -1.5 $      & 20       & 0.7 \\
Q16     &      $ -1.6 $      & 8.0      & 1.0 \\ 
Q17     &      $ -1.7 $      & 4.0      & 1.3 \\   
\bf{Q18}& $ \bf{-1.8} $      & \bf{2.0} & \bf{1.3} \\ 
Q19     &      $ -1.9 $      & 1.5      & 1.6 \\
Q20     &      $ -2.0 $      & 1.0      & 1.4 \\
\hline
\end{tabular}
\begin{flushleft}
\footnotesize{{\bf Notes:}}
\footnotesize{ Q18 is our fiducial model. }
\end{flushleft}
\label{t1}
\end{table}
%
%

Note that, for GeV $\gamma$-rays our constructed SED is not the intrinsic SED because we ignored the effect 
of the EBL on their propagation. These $\gamma$-ray photons while traveling through the IGM get annihilated 
upon collision with the EBL photons via electron-positron pair-production. This phenomena provides an effective 
optical depth $\tau_{\gamma}(\nu, z)$ for $\gamma$-rays which were emitted at redshift $z$ with frequency 
$\nu (1+z)$ and observed on earth with frequency $\nu$. Therefore, in our formalism the intrinsic (rest-frame) 
SED of $\gamma$-ray blazars at redshift $z$ should be $k^{\rm Q2}_{\nu} e^{\tau_{\gamma}(\nu (1+z), z)}$. In 
Appendix~\ref{app.gamma_tau}, we provide the values of $\tau_{\gamma}(\nu, z)$ from our EBL models which are 
mostly dominated by photons having energy lower than 10 eV and therefore insensitive to the value of assumed 
$\alpha$ or the QSO emissivity. 
Although we choose this parametric SED approach over standard X-ray population synthesis models, in Appendix~\ref{x-ray_emissivity} 
we show that this approach reproduces soft and hard X-ray emissivity and its redshift evolution reasonably well.

In the right-hand panel of Fig.~\ref{fig1}, we show our full composite QSO SED which is an addition of 
$k^{\rm Q1}_{\nu}$ and $k^{\rm Q2}_{\nu}$ obtained for $\alpha=-1.8$. To obtain the QSO emissivity at $z=0$ the 
full SED is normalized to the $\epsilon^Q_{\nu_{1000}}$ at $z=0$. As shown in the figure, the type-1 QSOs 
contribute mainly at $E<2$ keV, type-2 QSOs at $E>2$ keV till MeV and blazars at higher energies. At \hi~and 
\heii~ionizing energies, contribution from type-2 QSOs is extremely small. Therefore, 912 \AA~emissivity from 
all QSOs can be effectively written as 
%
\begin{equation}\label{eqso912}
\epsilon^Q_{912}(z)=\Big(\frac{1000}{912}\Big)^{\alpha}\epsilon^Q_{1000}(z)\,\,.
\end{equation}
%
%
One can interpret the ratio $k^{\rm Q2}_{\nu}/k_{\nu}$, which depends on choice of $\alpha$, as the fraction of 
type-2 QSOs (including highly obscured Compton thick QSOs and $\gamma$-ray blazars), however such an interpretation 
is subject to the shape of X-ray SED assumed for type-1 QSOs. We have further discussed such an interpretation in 
Appendix~\ref{app.fraction}. We do not model QSO SED in infra-red (IR). This does not affect our background 
calculations since QSOs contribute negligibly to IR background \citep[see also][]{Sazonov04}. Galaxies are more 
important in the optical and IR wavelengths. We discuss the galaxy emissivity in the following subsection.

%
\begin{figure*}
\centering
\includegraphics[totalheight=0.745\textheight, trim=0.0cm 0.2cm 7.5cm 0.0cm, clip=true, angle=270]{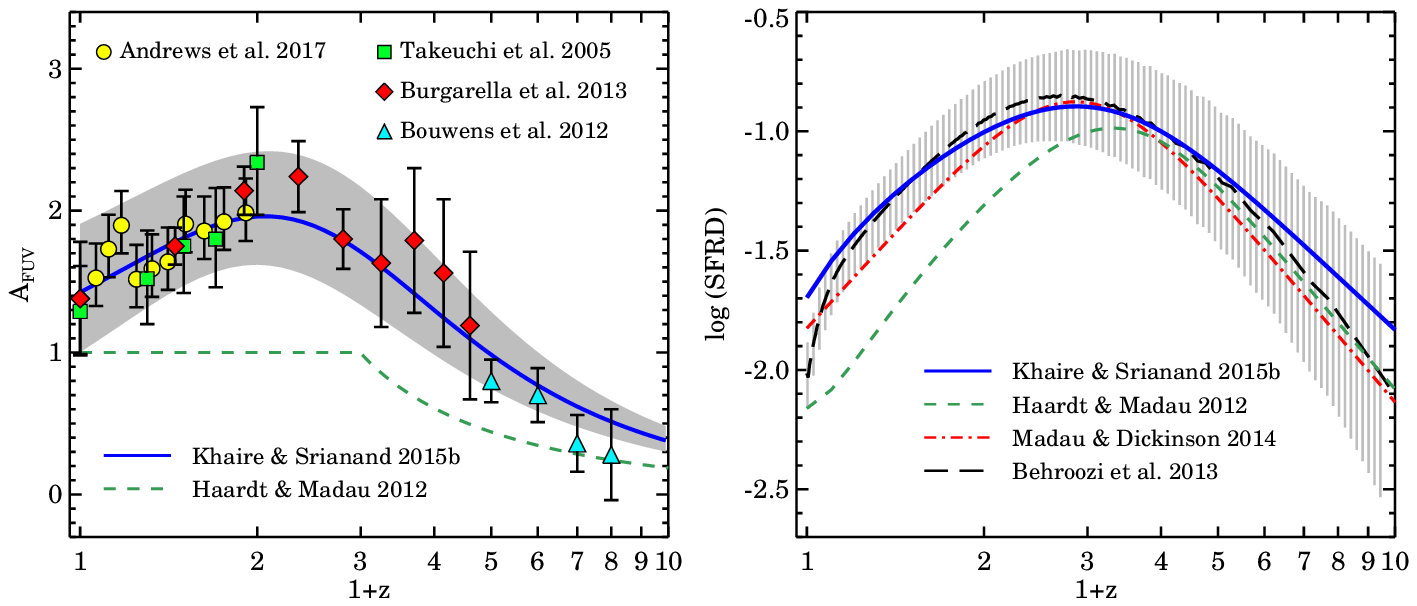}
\caption{
Left-hand panel: the average dust attenuation at FUV band ($A_{\rm FUV}$) 
in magnitudes with $z$. The solid-curve shows $A_{\rm FUV}(z)$ used in our 
EBL (equation~\ref{eq.afuv}) and the gray-shaded region shows the range in 
the $A_{\rm FUV}(z)$ arising from the scatter in the FUV luminosity 
functions of the galaxies \citepalias{Khaire15ebl}. Measurements by 
\citet[][circles]{Andrews17} are obtained by fitting SEDs of large number
of galaxies. Data points from \citet[][squares]{Takeuchi05} and 
\citet[][diamonds]{Burgarella13} were obtained using IRX-$\beta$ relation 
on IR and FUV luminosity functions of the galaxies. 
\citet[][triangles]{Bouwens12} measurements used the slopes of high-$z$ 
galaxy SEDs. We also show $A_{\rm FUV}(z)$ used by \citetalias{HM12} 
(dash curve) for comparison. 
Right-hand panel: the SFRD($z$) in units of ${\rm M_{\odot}\, yr^{-1}\, Mpc^{-3}}$. The solid-curve shows our fiducial 
SFRD($z$) obtained using the $A_{\rm FUV}(z)$ shown in the left-hand panel 
(equation~\ref{eq.sfrd}). For comparison we also show the SFRD estimated 
by \citet[][dot-dash curve]{Madau14}, \citet[][big-dash curve]{Behroozi13} 
and by \citetalias{HM12} (small-dash curve). The vertical-striped region 
shows the 1$\sigma$ uncertainty in the SFRD($z$) from \citet[][]{Behroozi13}. 
At $z<2$ \citetalias{HM12} SFRD($z$) is significantly different from others. 
}
\label{fig2}
\end{figure*}
%
%
\subsection{Galaxy Emissivity}\label{sec3.2}
The galaxy emissivity $\epsilon^G_{\nu}$ is obtained by integrating the galaxy luminosity functions (GLFs). To obtain
this at each $\nu$ from handful of GLF measurements, it is customary to use these GLFs to derive the star formation 
rate density (SFRD) first and then generate $\epsilon^G_{\nu}$ at each $\nu$ using stellar population synthesis 
models. However, most of the star formation tracers, especially the far-UV luminosity function which probes the most 
distant Universe, suffer from the unknown dust attenuation intrinsic to galaxies. The SFRD obtained using the 
far-UV GLFs is degenerate with the assumed amount of the dust attenuation in the FUV band (\afuv). We addressed this 
issue in \citetalias{Khaire15ebl}, where by using multi-wavelength GLFs we lifted the degeneracy between SFRD($z$) 
and \afuvz~for an assumed extinction curve. We found that for Large Magellanic Cloud Supershell (LMC2) extinction curve 
\citep[from][]{Gordon03}, our predicted \afuvz~is remarkably consistent with its measurements (see left-hand panel of 
Fig.~\ref{fig2}) obtained using IRX-$\beta$ relation on FUV and FIR GLFs \citep{Takeuchi05, Burgarella13}. The high-$z$ 
extrapolated part of our \afuvz~is also consistent with the \afuvz~measurements obtained from the UV slopes of galaxies 
upto $z\sim 7$ \citep{Bouwens12}. The \afuvz~is crucial not only to get the background in UV-optical wavelengths but 
also in FIR which is mainly dominated by the dust re-emission from galaxies. 

We use the \afuvz~magnitudes and SFRD($z$) from the \citetalias{Khaire15ebl} obtained for LMC2 extinction curve. 
These are
%
\begin{equation}\label{eq.afuv}
A_{\rm FUV}(z)=\frac{1.42+0.93z}{1+(z/2.08)^{2.2}}\,
\end{equation}
%
\begin{equation}\label{eq.sfrd}
{\rm SFRD}(z)=10^{-2}\times \frac{2.01+8.48z}{1+(z/2.5)^{3.09}} \,\, {\rm M_{\odot}\, yr^{-1}\, Mpc^{-3}} .
\end{equation}
%
To determine the \afuvz~and SFRD$(z)$ in \citetalias{Khaire15ebl}, we integrated the compiled multi-wavelength  
GLFs down to $0.01{\rm L_{\nu_0}^*}$ and used a stellar population synthesis code {\sc starburst99} \citep{Leitherer99}. 
It uses Salpeter initial mass function \citep[IMF;][]{Salpeter55} with exponent $-2.35$ and the stellar mass 
range from 0.1 to 100 M$_{\odot}$ with a constant metallicity of 0.4 times the solar value (i.e, $Z=0.008$).  

In the left-hand panel of Fig.~\ref{fig2}, we show our \afuvz~(solid-curve) along with their independent 
measurements obtained using the IRX-$\beta$ relation \citep{Takeuchi05, Burgarella13, Bouwens12}. Our \afuvz~is 
also remarkably consistent with the recent measurements by \citet{Andrews17} obtained by fitting SEDs to 
large number of galaxies observed in GAMA and COSMOS surveys \citep[see also,][]{Driver16dust}.  
The gray-shaded region shows the uncertainty in the obtained \afuvz~arising from the scatter in the
reported far-UV GLFs \citepalias[see][]{Khaire15ebl}. For comparison, we also show 
\afuvz~used in \citetalias{HM12} for their UVB calculations,
which is significantly smaller than the measurements and our values. 
The difference in the \afuvz~leads to different emissivities and SFRD($z$) which can severely affect  
the estimates of the EBL as well as the required $f_{\rm esc}$ to be consistent with \ghi~measurements.
Our SFRD$(z)$ obtained from the \afuvz~and stellar population synthesis models
is shown in the right-hand panel of Fig.~\ref{fig2}. For comparison we also 
show the SFRD$(z)$ obtained by \citetalias{HM12}, \citet{Madau14}
and \citet[][scaled by factor 1.7 to match the differences in the IMF used]{Behroozi13}. SFRD$(z)$ of \citetalias{HM12} is 
significantly smaller than others due to the lower \afuvz, as small as 
factor of $\sim 3$ at $z<2$. At $z<6$ our SFRD$(z)$ agrees well with those of \citet{Behroozi13} and \citet{Madau14} 
with differences smaller than $0.1$ and $0.2$ dex, respectively. At $z>6$ our SFRD(z) is more owing to higher \afuvz, however, 
within the 1-$\sigma$ uncertainty given by \citet{Behroozi13} as shown by the striped region in the 
right-hand panel of Fig.~\ref{fig2} and also consistent with 
the SFRD$(z)$ from very high-$z$ GLFs \citep{Oesch14, Bouwens15, McLeod15}. 

We obtain the $\epsilon^G_{\nu}(z)$, using \afuvz~and SFRD($z$) 
from equation~(\ref{eq.afuv}) and (\ref{eq.sfrd}) in the following convolution integral
%
\begin{equation}\label{eq.galemis}
\epsilon^G_{\nu} (z)=C_{\nu}(z)\int_{z}^{\infty}\frac{{\rm SFRD}(z')
\,\,l_{\nu}[t(z)-t(z'), Z]\,\,dz'}{(1+z')H(z')}\, ,
\end{equation}
%
where $l_{\nu}[t(z)-t(z'), Z]$ is a specific luminosity obtained from a simple stellar population in 
units of erg s$^{-1}$ Hz$^{-1}$ per unit mass of stars formed having metallicity $Z$ and age $t_0=t(z)-t(z')$.
The $C_{\nu}(z)$ is,
%
\begin{equation}\label{Eq.conv}
C_{\nu}(z)=
\begin{cases}
 10^{-0.4\,A_{\rm FUV}(z)\frac{D_{\nu}}{D_{\rm FUV}}}, & {\,\rm at \,\,} \lambda >912{\rm \,\AA} \\
 f_{\rm esc}(z), & {\,\rm at \,\,} 228 <\lambda < 912{\rm \,\AA} \\
 0 ,             & {\,\rm at \,\,} \lambda < 228{\rm \,\AA}.
\end{cases}
\end{equation}
%
At $\lambda>912$\AA, $C_{\nu}(z)$ is the dust correction where $D_{\nu}/D_{\rm FUV}$ is the LMC2 extinction curve 
$D_{\nu}$ normalized at FUV band.  
We take $C_{\nu}=f_{\rm esc}$ at $228 <\lambda < 912{\rm \,\AA}$ which assumes that
the ionizing photons are predominantly escaping through low-density low-dust 
channels in the galaxies \citep{Fujita03, Paardekooper11} or these photons are mostly generated by unobscured 
runaway stars in the outskirts of galaxies \citep{Gnedin08, Conroy12}. Note that we are treating $f_{\rm esc}$
in the same way as previous UVB calculations (such as \citetalias{HM12} and \citetalias{FG09}) treated.
We assume that no helium ionizing photons 
($\lambda < 228{\rm \,\AA}$) escape from galaxies. It is a reasonable assumption for galaxies at $z<6$ 
since the stellar population generates a negligibly small amount of high energy photons in the 
absence of a large number of population~{\sc iii} stars.

The $\epsilon^G_{\nu}(z)$ obtained by this method at $228<\lambda<912$ \AA~can be approximated as a power-law
\begin{equation}\label{eq.app}
\epsilon^G_{\nu}(z)= \epsilon^G_{\nu_{912}}(z) \Big(\frac{\nu}{\nu_{912}} \Big)^{\beta}\,.
\end{equation}
We find that $\beta=-1.8$ provides a best fit \citep[see also][]{Becker13}. We use this power-law approximation for 
simplicity (only for $\lambda <912$ \AA), to speed up the UVB calculations, and to smooth out the small fluctuations 
in $\epsilon^G_{\nu}$ produced by population synthesis models around $\sim$900 \AA.
We also verify that this power-law with $\beta=-1.8$ reproduces the number of photons
and the $\Gamma_{\rm HI}$ obtained by original $\epsilon^G_{\nu}$. 
A simple fit to the redshift evolution of our $\epsilon^G_{\nu_{912}}$ in units of ${\rm erg\, s^{-1}\, Hz^{-1} Mpc^{-3}}$ is given by
%
\begin{equation}\label{eq.gal_912}
\epsilon^G_{\nu_{912}}(z)=f_{\rm esc}(z) \times 10^{25} \frac{3.02+13.12z}{1 +(z/2.44)^{3.02}}\,.
\end{equation}
%
The extreme-UV emissivity from galaxies depends on the value of $f_{\rm esc}(z)$. Therefore, an accurate estimate of this is required to 
model the UVB.
%
%
\begin{figure*}
\centering
\includegraphics[totalheight=0.745\textheight, trim=0.0cm 0.1cm 7.5cm 0.0cm, clip=true, angle=270]{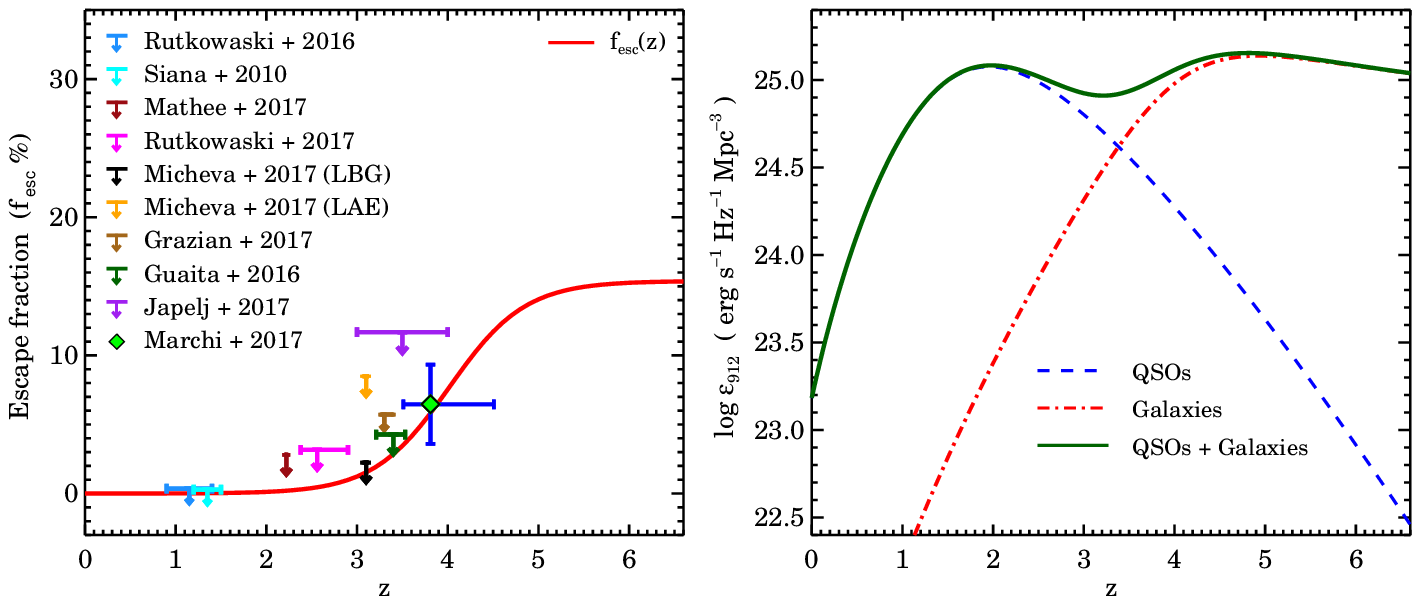}
\caption{
Left-hand panel: the average escape fraction $f_{\rm esc} (z)$ of \hi~ionizing 
photons from galaxies used in our UVB code (solid-curve, equation~\ref{eq.esc}).
Various data points show the recent estimates of average $f_{\rm esc}$ where 
downward-arrows indicate 1$\sigma$ upper limits. These points have been converted for 
our fiducial galaxy emissivity models as explained in the text and provided in 
Table~\ref{tab.escape}. Horizontal bars on each point indicate the redshift range of 
the galaxy sample considered for measurements.
Right-hand panel: the net specific emissivity at $912$\AA~($\epsilon_{\nu_{912}}$) 
with $z$ used in our fiducial UVB model (solid curve). The panel also shows the 
contribution from QSOs (dash-curve, see equation~\ref{eqso912} for $\alpha=-1.8$),
galaxies (dot-dash curve, from equation~\ref{eq.gal_912} and \ref{eq.esc} obtained 
for the $f_{\rm esc} (z)$ shown in the left-hand panel) and the addition of both 
(solid curve).  
}
\label{fig3}
\end{figure*}
%
%

\begin{figure*}
\centering
\includegraphics[totalheight=0.745\textheight, trim=7.5cm 0.0cm 0.1cm 0.0cm, clip=true, angle=90]{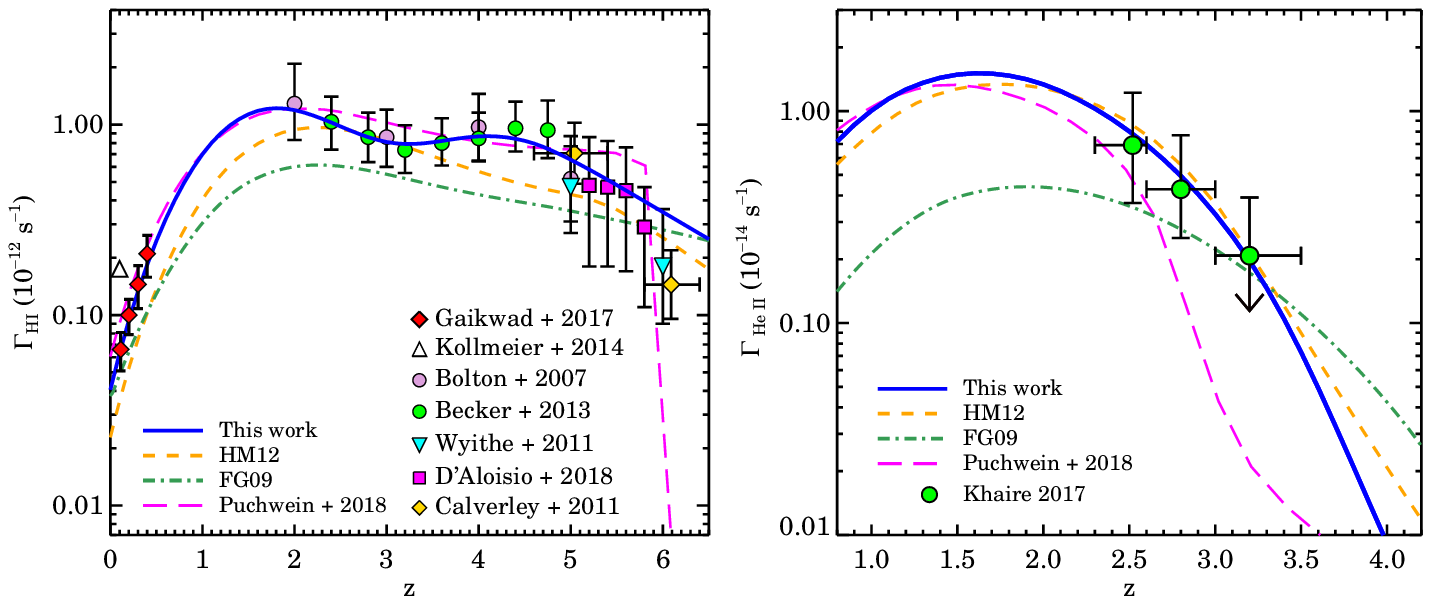}
\caption{
The photoionization rate of \hi~($\Gamma_{\rm HI}$; left-hand panel) 
and \heii~($\Gamma_{\rm He II}$; right-hand panel) as a function of $z$ 
from our fiducial UVB model (solid curves; see Fig.~\ref{figA1} for UVB 
models with different $\alpha$). For comparison we show the 
$\Gamma_{\rm HI}$ from UVB models of \citetalias{HM12} (small-dash curve) and
\citetalias{FG09} (dot-dash curve) 
and new model by \citet[][big-dash curve]{Puchwein18}. Various data points in the left-hand 
panel show the recent measurements of $\Gamma_{\rm HI}$. 
The $\Gamma_{\rm He II}$ from \citet{Khaire17sed} is obtained by using the 
measurements of $\tau_{\alpha}^{\rm He II}$ from \citet{Worseck16}.
}
\label{fig4}
\end{figure*}

As mentioned in Section~\ref{sec1}, only handful of galaxies are 
detected to show emission of extreme-UV photons in large sample of galaxies. 
Most of the average $f_{\rm esc}$ measurements are upper limits. 
Therefore, we resort to deduce the $f_{\rm esc}(z)$ that is required
to be consistent with the well-measured $\Gamma_{\rm HI}$ values as 
demonstrated in \citet{Khaire16}. We choose the $f_{\rm esc}(z)$ to be
%
\begin{equation}\label{eq.esc}
f_{\rm esc}(z)=\frac{10^{-5}}{6.5\times10^{-5}+\exp(-2.4z)}\,,
\end{equation}
%
%
which is consistent with the  $f_{\rm esc}(z)$ from \citet{Khaire16} and reproduce the $\Gamma_{\rm HI}(z)$ measurements 
(see Section~\ref{sec4.1} and Fig.~\ref{fig4}). 
In the left-hand panel of Fig.~\ref{fig3}, we compare our $f_{\rm esc}(z)$ with recent measurements 
from literature. The direct measurements of $f_{\rm esc}(z)$ are obtained using the observed ratio of flux ($f_{\lambda_{i}}$) at some 
extreme-UV wavelength
$\lambda_{i} \le 912$\AA~to the flux $f_{\lambda_{\rm FUV}}$ at some other higher wavelength, mostly in FUV band.
To get the $f_{\rm esc}(z)$ value, in addition to this observed ratio  $f_{\lambda_{i}}/f_{\lambda_{\rm FUV}}$, 
three other model dependent quantities are required.  
These are the intrinsic ratio $f_{\lambda_{i}}/f_{\lambda_{\rm FUV}}$ obtained from stellar population synthesis of the 
assumed galaxy model, the dust attenuation $A_{\lambda_{\rm FUV}}$ and the correction for the transmission
through IGM ($e^{\tau_{\rm eff}}$).  We have corrected the measurements shown in Fig.~\ref{fig3} for our values of $A_{\lambda_{\rm FUV}}$ and the
intrinsic ratio $f_{\lambda_{i}}/f_{\lambda_{\rm FUV}}$. For readers, these values are provided in Table~\ref{tab.escape}. 
We have taken the same $e^{\tau_{\rm eff}}$ used in the respective papers since, like us, 
most of them use the \hi~column-density distribution from \citet{InoueAK14}. 
As shown in Fig.~\ref{fig3}, our $f_{\rm esc}(z)$ is consistent with the measurements claimed by 
\citet{Marchi17} and all other $1\sigma$ upper 
limits \citep{Siana10, Guaita16, Rutkowski16, Rutkowski17, Matthee17, Micheva17, Grazian17, Japelj17}. 
Following \citet{Khaire16}, in absence of any observational evidence for the evolution of  
$f_{\rm esc}$ at $z>6$, our $f_{\rm esc}(z)$ reaches a constant asymptotic value of $~0.15$. 
As shown later, this $f_{\rm esc}$ value gives \hi~reionization history consistent with various observations. 

In the right-hand panel of Fig.~\ref{fig3}, we show the relative 
contribution of galaxies and QSOs to the extreme-UV emissivities, in terms of
$\epsilon_{912}(z)$. We show QSO contribution 
$\epsilon^Q_{912}(z)$ for $\alpha=-1.8$ and our galaxy contribution $\epsilon^G_{912}(z)$ 
(from equation~\ref{eq.gal_912} and \ref{eq.esc}). QSOs dominate the extreme-UV emissivity 
at $z<3$ (irrespective of the assumed $\alpha$ value) and galaxies 
dominate at $z>3.5$. The sharp increase and then saturation in $\epsilon^G_{912}(z)$ results from the 
$f_{\rm esc}(z)$ adopted in our study (see the left hand panel of Fig.~\ref{fig3}). 
The rapid decrease in the $\epsilon^Q_{912}(z)$ at $z>3$
is the main reason for the required rapid increase in $f_{\rm esc}(z)$ and hence in 
the $\epsilon^G_{912}(z)$ at $3.5<z<5.5$ in order to
be consistent with the $\Gamma_{\rm HI}$ measurements \citep{Khaire16}. 
If low luminosity AGNs contribute
appreciably to the emissivity
at $z>3.5$, as claimed by
\citet{Giallongo15}, one may not need such a rapid increase in $f_{\rm esc}(z)$ \citep[see e.g.,][]{Khaire16}. We discuss 
the effect on UVB arising from uncertainties in galaxy emissivity in Section~\ref{sec4.4.2}.

The galaxy emissivity obtained using equation~(\ref{eq.galemis}) does not provide the IR and FIR emissivities since it 
includes only stellar radiation. Radiation from old stellar population peaks in near-IR around 1-3$\mu$m and falls 
steeply at smaller wavelengths. Most of the observed FIR emission from galaxies originate from the thermal emission 
of interstellar dust heated by UV and optical light from stars. We take the FIR emissivity estimated by 
\citetalias{Khaire15ebl} for the same
SFRD$(z)$, $A_{\rm FUV}(z)$ and the LMC2 extinction curve used here. It has been estimated under the assumption that the 
energy density absorbed by dust in UV and optical wavelengths is getting emitted in IR to FIR wavelengths and 
the spectral shape of this emission is similar to the one observed from galaxies in local Universe. 
Such a spectral shape has been taken 
from local IR galaxy templates of \citet{Rieke09}. This FIR emissivity has been shown to be consistent with various local 
observations. For more details on this we refer readers to the section 5 of \citetalias{Khaire15ebl}.

\subsection{Diffuse Emissivity}\label{sec3.3}
Most of the gas in the IGM is in photoionization equilibrium with the UVB.  
This gas re-emits a fraction of energy it absorbs from the UVB at 
different wavelengths. 
This re-emission happens through various recombination channels such 
as Lyman-series and Lyman-continuum emission of \hi, \hei~and \heii, similarly for 
Balmer and higher order series and continuum. Although, it contributes negligibly to 
UVB as shown by \citetalias{FG09}, we model few of the most dominant contributions in the  
extreme-UV wavelengths such as Lyman-continuum emission from \hi~and \heii, Balmer-continuum
emission from \heii~and the Lyman-$\alpha$ emission from \heii~following the procedure in \citetalias{FG09} and \citetalias{HM12}. 
We briefly describe it here. For more details we refer readers to the relevant 
sections in \citetalias{FG09} and \citetalias{HM12}.

For diffuse emission from \hi~and \heii~we follow analytic approximations given in \citetalias{FG09}. The 
comoving recombination emissivity is obtained by solving following equation
\begin{equation}\label{eq.recomb}
\epsilon^{\rm d}_{\nu}(z)=\frac{4\pi H(z)}{(1+z)^2}  \int_0^{\infty} dN_{\rm HI} f(N_{\rm HI}, z) I_{\nu}^{\rm rec} (N_{\rm HI}),
\end{equation}
%
where 
\begin{equation}\label{eq.recombI}
I_{\nu}^{\rm rec} (N_x)=\frac{h\nu_{\rm rec}}{4\pi} \frac{\alpha_{\rm rec}(\rm T)}{\alpha^A_{\rm x}(\rm T)}N_{x,\rm th}(1-e^{-N_{x}/N_{x, \rm th}}) \Gamma_{x} \phi^{\rm rec}_{\nu}.
\end{equation}
%
Here, the subscript $x$ denotes species (i.e \hi~or~\heii), $\nu_{\rm rec}$ is the line frequency of the recombination emission, $\alpha_{\rm rec}$ is
the recombination rate coefficient for the relevant transition and $\phi^{\rm rec}_{\nu}$ is 
the recombination line profile. Following \citetalias{FG09}, for \hi~and \heii~Lyman continuum emission
we use $N_{\rm HI, th}=10^{16.75}$ cm$^{-2}$ and $N_{\rm HeII, th}=10^{17.3}$ cm$^{-2}$, 
and for \heii~Balmer continuum emission and  \heii~Lyman-$\alpha$ line radiation 
we use $N_{\rm He II, th}=2.3 \times 10^{18}$ cm$^{-2}$ following \citetalias{HM12}.
For the continuum emission we use line profile
\begin{equation}\label{eq.recomb}
\phi^{\rm rec}_{\nu}= \frac{(\nu/\nu_{\rm rec})^{-1} \exp(h\nu /k\rm T)}{\Gamma(0, h\nu_{\rm rec} /k\rm T)} \, \frac{\Theta(\nu -\nu_{\rm rec})}{\nu_{\rm rec}},
\end{equation}
%
where $\Theta$ is the Heaviside function and $\Gamma(0, h\nu_{\rm rec} /k\rm T)$ is incomplete gamma function. For \heii~Lyman-$\alpha$ line emission, 
we use $\phi^{\rm rec}_{\nu}=\delta (\nu - \nu_{\alpha})$ where $\delta$ represents Dirac-delta function and $\nu_{\alpha}=c/303.78\,{\rm \AA}$. We do not model the \hi~Lyman-$\alpha$ line emission from IGM. 
The contribution of \hi~Lyman-$\alpha$ line emissivity to the total emissivity is negligible because 
the emissivity from galaxies at $1216 \,\rm \AA$ is more than an order of magnitude higher \citepalias[see][Fig.10]{HM12}. For all calculations we assume that the IGM
is at $\rm T=20000$ K. 


We have also modeled the \hi~Lyman-$\alpha$ emissivity from galaxies which is not included in the population synthesis models. 
For this we followed the simple procedure used by \citetalias{HM12} (their section 7.1), which assumes that
68\% of the Lyman continuum photons which are not able escape the galaxies (i.e, $(1-f_{\rm esc})\times0.68$ ) 
are being emitted as Lyman-$\alpha$ photons. Unlike \citetalias{HM12}, we multiply this
\hi~Lyman-$\alpha$ emissivity by $10^{-0.4A_{\rm FUV}(z)D_{Ly\alpha}/D_{\rm FUV}}$ to capture the effect of the dust attenuation. 
Therefore, our UVB does not show prominent Lyman-$\alpha$ emission as compared to \citetalias{HM12}. 
Note that such a simple
method to estimate the \hi~Lyman-$\alpha$ emissivity may not be consistent with the complex radiative transfer and escape of Lyman-$\alpha$
photons emitted by stellar population through ISM and CGM of galaxies 
\citep[see e.g.,][]{Neufeld90, Neufeld91, Dijkstra06, Dayal10, Gronke16, Dijkstra17}. 
%
%
%

\section{Basic Results}\label{sec4}
In this section we discuss basic results from our fiducial model (Q18 model with $\alpha=-1.8$ used in QSO SED) 
for the photoionization and photoheating rates of \hi~and \heii, the reionization history of \hi~and \heii, the spectrum of local 
EBL from FIR to $\gamma$-rays and the UVB spectrum along with its comparison with models of \citetalias{FG09} and \citetalias{HM12}.
Similar results for the models obtained with different $\alpha$ are presented in the Appendix~\ref{app.other_uvbs}. 
We also discuss the uncertainties in the UVB arising from different model parameters. 

%
\subsection{Photoionization and photoheating rates}\label{sec4.1}
The photoionization rates of \hi~and \heii~are the only observed constraints on the UVB. Moreover, 
these are just integral constraints and can decide the intensity of the UVB if only the spectral shape of 
the UVB is known. However, unlike the direct observational constraints on EBL at wavelengths other than extreme-UV, 
photoionization rates are not limited to the local Universe and can be measured across a large redshift range 
using various observational techniques as mentioned in Section~\ref{sec1}.  
In our analysis, we use the recent \ghi~measurements obtained by comparing the cosmological simulations of the IGM to the 
observed statistics of the Ly-$\alpha$ forest by \citet{Gaikwad17a} at $z<0.5$, by \citet{Becker13} and \citet{Bolton07}
at $2<z<5$, and by \citet{DAloisio18} and \citet{Wyithe11} at $5<z<6$. 
We also use the \ghi~measurements obtained from proximity zones of high-$z$
QSOs by \citet{Calverley11} at $z=5$ and 6.1. 
In addition to \ghi, we have also used the \gheii~measurements by \citet{Khaire17sed}.

In the left-hand panel of the Fig.~\ref{fig4}, we show the \ghiz~obtained 
from our fiducial UVB model along with various available measurements. The 
remarkable agreement between our prediction and the \ghi~measurements at $z>3$ is not surprising since we have chosen the 
appropriate \fescz~to match these measurements \citep[see][]{Khaire16}. Whereas, at $z<3$ the UVB is determined by our 
updated QSO emissivity alone \citepalias[see also][]{Khaire15puc} and there is no need to adjust \fescz. However, for making 
\fescz~as a continuous function of $z$, we have taken negligibly small values of \fescz\; at $z<2.5$. 
At $z<0.5$, our \ghi~is not only consistent with measurements from \citet{Gaikwad17a} but
also with \citet{Shull15, Gurvich17, Viel17, Khaire18ps} and \citet{Fumagalli17}, which are not shown in the figure due to shortage of space. 
The $z=0$ \citet{Fumagalli17} measurement is obtained using the observations of
H$\alpha$ fluorescence from a nearby faint disc galaxy. 
Note that all these $z<0.5$ measurements provide $\sim$2.5 times smaller \ghi~than the measurements of \citet{Kollmeier14}. 
For comparison, we also show the \ghiz~obtained in UVB models of \citetalias{HM12}, \citetalias{FG09}\footnote{We
use their December 2011 update available on web-page http://galaxies.northwestern.edu/}
and a new model of \citet{Puchwein18}.
At $z<0.5$, unlike our models 
previous two models by \citetalias{HM12} and \citetalias{FG09}
under-predict \ghi~values compared to latest measurements.
This is one of our major improvements over these previous
UVB models. The differences in the \ghiz~predicted by these models at $z>2$ are because of 
choosing different inputs in their UVB models, such as
$f_{\rm esc}$, in order to be consistent with
different \ghiz~measurements. For example, \citetalias{FG09} tried to 
be consistent with  \ghiz~from \citet{FG08} whereas \citetalias{HM12} with
\ghiz~from \citet{Becker07}, which are quite different from the recent measurements \citep[see][for more details]{Becker13}.
Overall, our \ghiz~values are in excellent agreement with the recent measurements at all redshifts than previous UVB models. 
The \ghi~predicted by the new UVB model of \citet{Puchwein18} is also in very good agreement with recent measurements and our model predictions at $z<5.5$. 
This new model has employed a novel opacity treatment 
to synthesize UVB in the pre-\hi~and \heii-reionization era following \citet{Madau17}. A sharply decreasing \ghiz~at $z>5.8$ is one of the byproducts of such opacity
treatment which has been shown to be important in reproducing reionization history and heating in hydrodynamical simulations of the IGM \citep[see also][]{Jose17}. 
However, \citet{Puchwein18} success at $z<1$ to reproduce recent \ghi~measurements mainly 
comes from using updated QSO emissivity similar to our model \citep[see also][]{Madau15}. 
Note that, the uncertainties in the \ghi~measurements are different for different measurement methods. Even for the same
method, such as using flux decrement, the uncertainties depend on choice of range in the thermal state of the IGM used in calculation \citep[see][]{DAloisio18}. 
Therefore, the agreement between UVB models and the \ghi~measurements also depend on the \ghi~measurements considered in the model.

UVB models with different $\alpha$ give slightly different values of \ghi~at $z<3$ because
at these redshifts the UVB is dominated by QSOs. However, these are still
consistent with the measurements of \citet{Gaikwad17a} as shown in the left-hand panel of Fig.~\ref{figA1} 
in the Appendix. This shows that the 
difference in the obtained \ghi~due to changing $\alpha$ (from -1.4 to -2) is smaller than 
the present uncertainties on the low-$z$ \ghi~measurements. 
At $z>3$ the \ghi~is dominated by galaxy emissivity
therefore left-hand panel of Fig.~\ref{figA1} does not show any change in \ghi~with $\alpha$.  

In the right-hand panel of the Fig.~\ref{fig4}, we show the \gheiiz~obtained from our fiducial UVB  
along with the values determined 
in \citet{Khaire17sed} at $2.5<z<3.5$ by using the \heii~effective optical depth measurements of \citet{Worseck16} 
and the $f(N_{\rm HI}, z)$ from \citet{InoueAK14}. 
For comparison, we also show the \gheiiz~from UVB models 
of \citetalias{HM12}, \citetalias{FG09}
and a new model of \citet{Puchwein18}.
\gheiiz~obtained by our and \citetalias{HM12} UVB models 
match very well with the measurements. 
In this $z$-range where direct observations are available, the \gheiiz~from \citetalias{FG09}
and \citet{Puchwein18} are also broadly consistent with \citet{Khaire17sed} measurements.
The agreement between our and \citetalias{HM12} \gheii~at $2.5<z<3.5$, despite the fact that we are using different 
input parameters, is mostly coincidental and because of using different QSO SEDs. 
\citetalias{HM12} used smaller $\epsilon^Q_{1000}$ but steeper SED ($\alpha=-1.57$) 
on the other hand we used higher $\epsilon^Q_{1000}$ but
shallower SED ($\alpha=-1.8$). Also, the fact that both models get almost the 
same \ghi~within $2.5<z<3.5$ helps in getting the \gheiiz~agreement.

The \gheii~is more sensitive to the  value of $\alpha$ than \ghi. 
It is because $\alpha$ is obtained by normalizing QSO emissivity at $\lambda = 1000$ \AA. 
Therefore, a small change in $\alpha$ results in a large change for 
\heii~ionizing photons at more than four times smaller wavelengths. 
We show \gheii~obtained from our UVB models with different $\alpha$ 
in the right-hand panel of Fig.~\ref{figA2}. 
Only $-2.0<\alpha<-1.6$ are consistent with the \gheii~measurements and 
\heii~effective optical depths \citep[for more detailed 
analysis on this refer to][]{Khaire17sed}.
We note that new measurement of \gheiiz~by \citet{Worseck18} are consistent with our UVB predictions 
for $\alpha=-2.0$.

In Table~\ref{ta2} we provide the photoionization and photoheating rates for 
\hi, \hei~and \heii~from our fiducial UVB model with $\alpha=-1.8$
and in Tables~\ref{ta3}-\ref{ta8} for UVB models with different $\alpha$. 
Photoheating rates for different species $x$ are obtained by
%
\begin{equation}\label{eq.xi}
{\xi}_{x}(z)=\int_{\nu_{x}}^{\infty}d\nu\,
\frac{4\pi\,J_{\nu}(z)}{h\nu}\,h(\nu-\nu_x)\sigma_{x}(\nu)\,\,.
\end{equation}
%
In Fig.~\ref{figA2} we show the 
$\xi_{\rm H I}(z)$ and $\xi_{\rm He II}(z)$ for our UVB models obtained with different values of $\alpha$.
The photoheating rates follow broadly the similar redshift evolution as photoionization rates (compare Fig.~\ref{figA1} and \ref{figA2}). 
As in the case for \gheii, the $\xi_{\rm He II}$ is also sensitive to the value of $\alpha$.  Note, by construct, all these models
consistently reproduce \ghiz~but differ only in \gheiiz.

\subsection{\hi~and \heii~reionization}\label{sec4.2}
%
\begin{figure*}
\centering
\includegraphics[totalheight=0.745\textheight, trim=0.0cm 0.1cm 11.0cm 0.0cm, clip=true, angle=270]{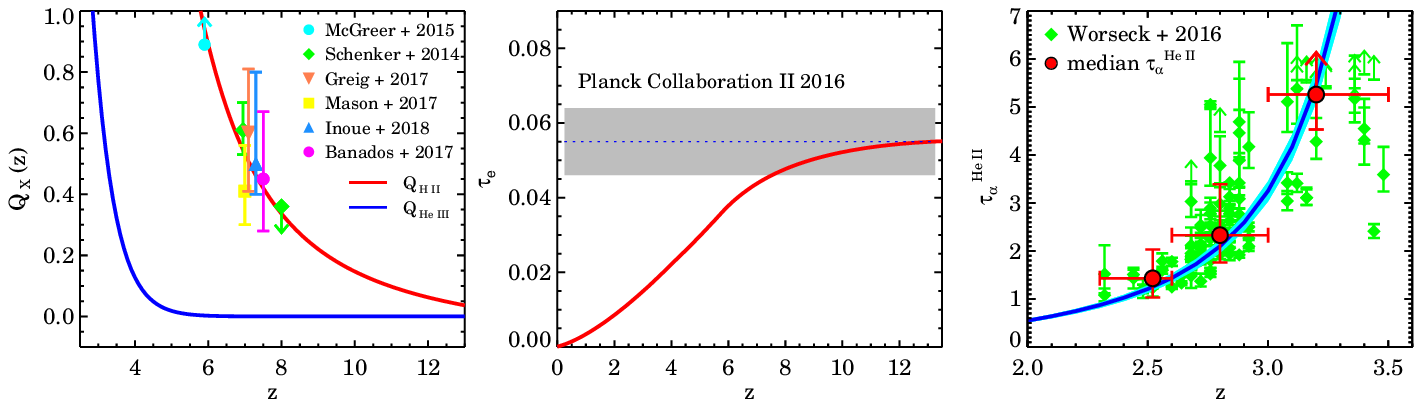}
\caption{\emph{Left-hand panel}: volume filling fraction of \hii~(red curve) and \heiii~(blue curve) obtained for our fiducial UVB model. 
For comparison the $Q_{\rm HII}$ measurements from \citet{Schenker14, Mcgreer15, Greig17, Mason17, Banados17, Inoue18} has been shown. 
We get $z_{\rm re}=5.8$ for  \hi~and $z_{\rm re}=2.8$ for \heii.
\emph{Central panel}: 
electron scattering optical depth ($\tau_e$) obtained from our fiducial UVB model (solid curve) shown together with the measurements from
\citet[][dotted line with gray shade]{Planck16b}.
\emph{Right-hand panel}: 
the \heii~Lyman-$\alpha$ effective optical depth (\theii) obtained from our 
fiducial UVB model (solid curves with $b=28$ km s$^{-1}$ and cyan shade obtained by changing $b$ from 24 to 32 km s$^{-1}$) 
with the measurements from \citet[][diamonds]{Worseck16}. The red points show the median 
$\tau_{\alpha}^{\rm He II}$ in redshift bins indicated by horizontal bars. 
}
\label{fig5}
\end{figure*}
%
%
We have calculated the \hi~and \heii~reionization history for the \hi~and \heii~ionizing emissivity used in our fiducial UVB model. The redshift evolution of
volume filling factor $Q_x$ of \hii~and \heiii~are obtained by solving \citep[see][]{Madau99, Khaire16},
%
\begin{equation}\label{q}
\begin{aligned}
Q_x(z_0)=\frac{1}{\langle n \rangle} \int^{\infty}_{z_0} dz\,\frac{\dot n_y(z)}{(1+z)H(z)}\,\times\\ 
\exp \Bigg[-\alpha^{\rm B}_y(T) \langle n \rangle 
 \int^{z}_{z_0}{dz'\frac{\chi(z') C(z')(1+z')^2}{H(z')}}\Bigg] \,\,.
\end{aligned}
\end{equation} 
%
where, $\langle n  \rangle=\langle n_{\rm H} \rangle$ and $y=$ \hi~when $x$ is 
\hii~and  $\langle n \rangle=\langle n_{\rm He} \rangle$ 
and $y=$ \heii~when $x$ is \heiii,  $\dot n_y(t)$ is the comoving number 
density of ionizing photons per unit time for species $y$, $C(z)$ is the 
clumping factor and $\chi(z)$ is number of photoelectrons per
hydrogen atom at redshift $z$. We use $\chi(z)=1.083$ at $z>4$ and $\chi(z)=1.16$ at $z<4$ assuming 
that within the regions where \hi~is ionized the  
helium is predominantly singly ionized at $z>4$ and doubly ionized at $z<4$.  
We use the comoving number density of hydrogen $\langle n_{\rm H} \rangle=1.87\times10^{-7} $ cm$^{-3}$ and helium 
$\langle n_{\rm He} \rangle= \langle n_{\rm H} \rangle y_p/(4-4y_p)$ where 
$y_p=0.248$ is the helium mass fraction from \citet{Planck16}.
The $\dot n_y(z)$ is obtained by
%
\begin{equation}\label{ndot}
\dot n_y(z) = \int^{\infty}_{\nu_y}d\nu\,{\frac{\epsilon_{\nu} (z)}{h \nu}}\,\,,
\end{equation} 
%
where ${\nu}_y$ is threshold ionization frequency for species $y$.
We use clumping factor $C(z)=9.25-7.21 \log(1+z)$ obtained from the cosmological hydrodynamical simulations of the IGM 
from \citet{Finlator12}. 
%
%
\begin{figure*}
\centering
\includegraphics[totalheight=0.745\textheight, trim=0.0cm 0.1cm 5.2cm 0.0cm, clip=true, angle=270]{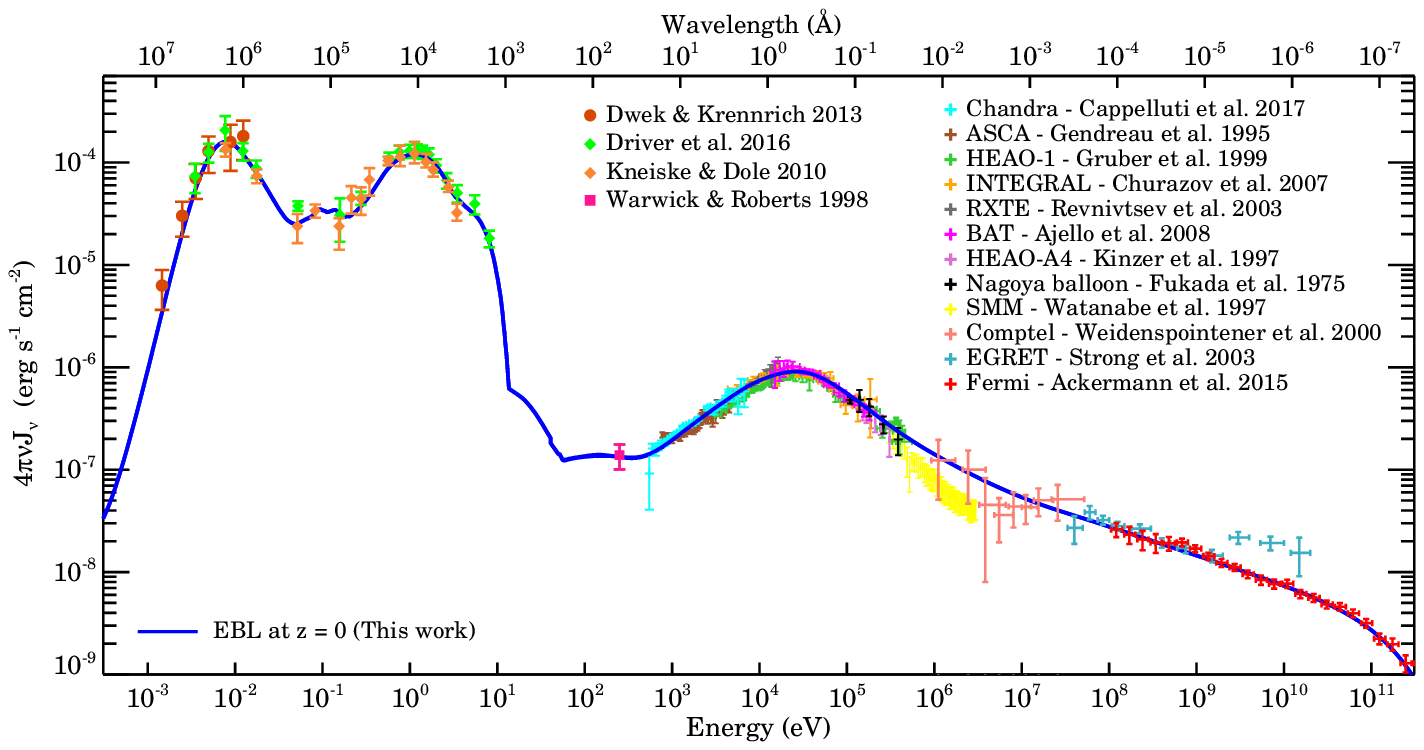}
\caption{The intensity of $z=0$ EBL from FIR to TeV $\gamma$-rays from our fiducial model (see Fig.~\ref{figA3} for EBL 
models with different $\alpha$). Various measurements shown in the 
figure are summarized in Section~\ref{sec4.3}.     
}
\label{fig6}
\end{figure*}
%
%

We show the $Q_{\rm H II}(z)$  and $Q_{\rm He III}(z)$ results in the left-hand panel of Fig.~\ref{fig5}. 
We also show recent measurements of $Q_{\rm HII}$. \hi~reionization in our model completes at $z=5.8$ ($z$ at which $Q_x(z)=1$) 
consistent with high-$z$ Lyman-$\alpha$ forest observations \citep{Beckerr01, Fan06}. Our $Q_{\rm H II}(z)$ 
is in agreement with lower limits from \citet{Mcgreer15} obtained using dark-pixel statistics of high-z QSO spectra and 
with measurements of \citet{Greig17} and \citet{Banados17} obtained using Lyman-$\alpha$ 
damping wings of two highest redshift QSOs.
Our $Q_{\rm H II}(z)$ is also consistent with the  measurements obtained from the diminishing population of high-$z$ 
Lyman-$\alpha$ emitters obtained by \citet{Schenker14}, \citet{Mason17} and \citet{Inoue18}. 
We calculate the electron scattering optical depth $\tau_e$ of CMB \citep[Eq.~12 from][]{Khaire16}.
The $\tau_e(z)$ from our reionization model is shown in the central panel of Fig.~\ref{fig5}, which 
reaches asymptotic value $\tau_e=0.55$ consistent with the measurements of \citet{Planck16b}. 
This reiterated the results of \citet{Khaire16} that new constraints on \hi~reionization
does not require  $f_{\rm esc}(z)$ at $z>6$ to have a steep evolution and a constant value, such as 0.15 used here 
(equation~\ref{eq.esc}), is sufficient to produce the consistent $\tau_e$. 
The \hi~reionization history is same for all of our UVB models
obtained with different $\alpha$ because in these models the 
\hi~reionization is driven by galaxies and QSOs contribute negligibly.

\heii~reionization in our fiducial model with $\alpha=-1.8$ completes at $z=2.8$. It is 
consistent with various measurements of \heii~Lyman-$\alpha$ 
effective optical depths \citep[\theii;][]{Kriss01, Shull04, Shull10, Fechner06, Worseck11, Worseck16}, 
measurements of peak in the redshift evolution of mean IGM temperature  
\citep{Becker11t, Hiss17} and also with theoretical models of \heii~reionization \citep{McQuinn09, Compostella13, Plante16}. 
We also consistently reproduce \theii~from \citet{Worseck16}, as shown in the right-hand panel of Fig.~\ref{fig5}. 
It shows \theii~calculated from our fiducial UVB model \citep[using 
equation 9 and 10 from][]{Khaire17sed} for an assumed Doppler 
broadening $b=28$ km/s with blue curve and the shaded cyan region provides
values encompassed by changing b from 24 to 32 km/s. The red data points, shown to guide the eyes, are 
the median values of \theii~(within 95 percentile errors) as obtained in three different $z$ bins
\citep[see table 2 of][]{Khaire17sed}. Note that the \heii~reionization history and \theii($z$) depends on the value of $\alpha$.
As shown in \citet{Khaire17sed}, only $-2.0<\alpha<-1.6$ are consistent with 
both \theii~measurements and epoch of \heii~reionization
being $2.6<z<3.0$.
Similar conclusions are presented in recent study by \citet{Gaikwad18} using the UVB models presented here in hydrodynamic simulations of the 
IGM explicitly taking into account the non-equilibrium ionization effects.

\subsection{The EBL spectrum}\label{sec4.3}
In Fig.~\ref{fig6}, we show our full EBL spectrum at $z=0$. Our calculations cover more than 
fifteen orders of magnitude range in wavelength 
from FIR to TeV energy $\gamma$-rays. The three distinct peaks in the intensity, 
$4 \pi \nu J_{\nu}$ in units of erg s$^{-1}$ cm$^{-2}$, can be readily seen. 
These are the FIR peak arising from the dust emission around 
$\sim$100 $\mu$m (10$^6$ \AA), the near IR peak dominated by old stellar population around 
$\sim$1 $\mu$m (10$^4$ \AA) and the hard X-ray peak from type-2 QSOs around $\sim 30$ keV ($0.4$ \AA).  
In general, the $\lambda>912$ \AA~part of the EBL is dominated by
emission from galaxies including their stellar and dust emission, the $\lambda<228$ \AA~part is contributed by radiation only 
from QSOs, and the $912>\lambda>228$ \AA~part is contributed by both galaxies and QSOs. 
Relative contribution to the latter is determined by $f_{\rm esc}$, which at 
$z=0$ is negligibly small therefore it is contributed by QSOs alone. 

In Fig.~\ref{fig6}, we also show $z=0$ EBL measurements at different wavelengths obtained by various methods 
and instruments. In the FIR wavelengths we use compiled measurements by \citet[][their table 7]{Dwek13} obtained from COBE. In  
the optical to FIR wavelengths we use compiled measurements by \citet{Kneiske10} and \citet{Driver16ebl} obtained from integrated
light from resolved galaxies in deep surveys. Our EBL models agree very well with these measurements.
At 0.25 keV ($\lambda \sim 50$ \AA) the measurement is taken from
\citet{Warwick98} obtained using shadow measurements from ROSAT. The X-ray background 
measurements are taken from various instruments;
Chandra \citep{Cappelluti17}, ASCA \citep{Gendreau95}, HEAO-1 and HEAO-4 \citep{Gruber99, Kinzer97},
Integral \citep{Churazov07}, RXTE \citep{Revnivtsev03}, Swift/BAT \citep{Ajello08}
and Nagoya balloon \citep{Fukada75}. Similarly $\gamma$-ray background measurements are taken from instruments:
SMM \citep{Watanabe97}, compton/Comptel \citep{Weidenspointner00}, compton/EGRET \citep{Strong03} and 
Fermi/LAT \citep{Ackermann15}. Our local fiducial EBL shows remarkable match with the X-ray and $\gamma$-ray background 
measurements all the way upto TeV $\gamma$-rays. It is not surprising, since we have constructed type-2 QSO and blazar SED 
(Section~\ref{sec3.1}) and adjusted the normalizations (see Table~\ref{t1}) to do so. 
In Fig.~\ref{figA3}, we show our EBL models with different values of $\alpha$.
The normalization $h \nu_0$ and $S_{k}$ are taken such that 
despite different $\alpha$, the local EBL will consistently reproduce the X-ray and $\gamma$-ray measurements 
as shown in the Fig.~\ref{figA3}. However, with our assumed QSO SED shape it becomes difficult for models with $\alpha > -1.5$, 
to match few of the soft X-ray measurements.

The optical depth encountered by high-energy $\gamma$-rays ($\tau_{\gamma}$) due to 
EBL are discussed in Appendix~\ref{app.gamma_tau}. We find that the 
$\tau_{\gamma}$ is insensitive to EBL at ${\rm E} < 10$ eV and thus to values of 
$\alpha$. Note that, the EBL at ${\rm E} < 10$ eV is same as the fiducial model
given in \citetalias{Khaire15ebl} (the median LMC2 model) 
and therefore it is consistent with many published models of optical and FIR EBL
\citep[e.g.,][]{Inoue13, Gilmore12, Finke10, Kneiske10, Franceschini08, Dominguez11, Helgason12, Scully14} 
as shown in the \citetalias{Khaire15ebl}. We provide the updated 
$\tau_{\gamma}$ values calculated for our fiducial EBL model including CMB.
%
\begin{figure*}
\centering
\includegraphics[totalheight=0.745\textheight, trim=0.0cm 0.1cm 2.8cm 0.0cm, clip=true, angle=270]{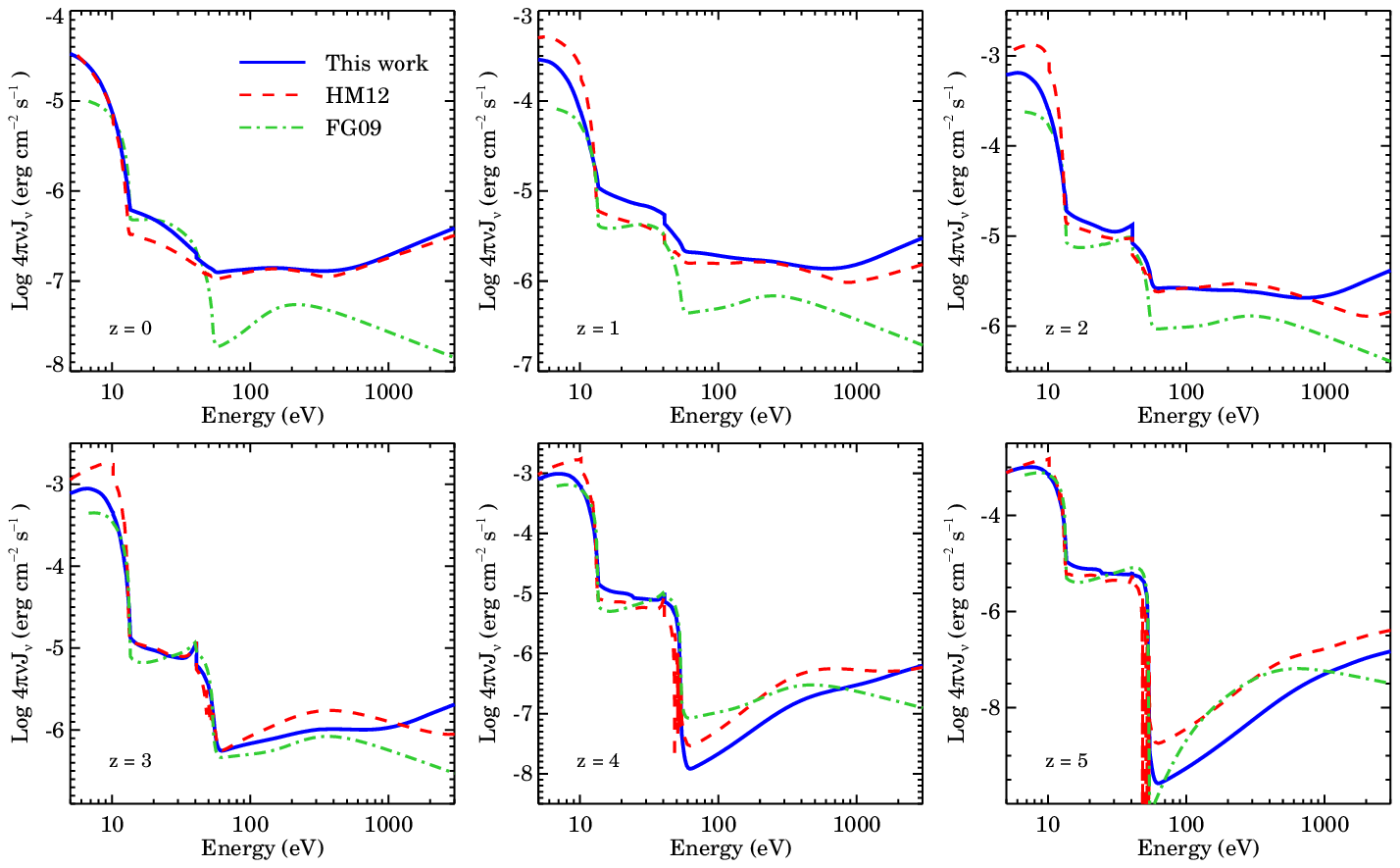}
\caption{The spectrum of our fiducial UVB with energy (from 5 to 3000 eV) at $z=0$, 1, 2, 3, 4 and 5. 
The solid blue curve shows our fiducial UVB model (see Fig.~\ref{figA4} for UVB 
models with different $\alpha$). The red dash and green dot-dash 
curves show the UVB models from \citetalias{HM12} and \citetalias{FG09}.}
\label{fig7}
\end{figure*}
%

The UVB at z=0, a part of the EBL at $ 50 <\lambda < 1200$ \AA~(i.e., $10<{\rm E}<250$ eV) 
shown in Fig.~\ref{fig6}, does not have any
direct measurements.  
The measurements such as photoionization rates provide only integral 
constraints. Our models are consistent with them across a large
$z$-range as shown in Fig.~\ref{fig4}. Below we discuss the UVB spectrum and its redshift evolution in 
comparison with the previous UVB models. For the discussions hereafter, 
let us naively divide the EBL at $\lambda < 912$~\AA~(${\rm E}>13.6$ eV) into three parts, the
\hi~ionizing UVB at $13.6<{\rm E}<54.4$ eV, the \heii~ionizing UVB at $54.4<{\rm E}<500$ eV and the
X-ray background at ${\rm E}>500$ eV.

In Fig.~\ref{fig7}, we compare our fiducial UVB spectrum from $z=0$ 
to 5 with the UVB from \citetalias{FG09} and \citetalias{HM12}. 
Two distinct features in the UVB spectrum can be readily seen, these are ionization edges of 
\hi~(at 13.6 eV) and \heii~(at 54.4 eV).
The smoothness of these features at the bottom of the troughs is because of the inclusion of Lyman continuum emission
from \hi~and \heii~in UVB calculations. Also there is 
a small kink at 40.8 eV, prominent at $z=1$ to 4 UVB, due to \heii~Lyman-$\alpha$ 
recombination emission. Another such small kink due to \hi~Lyman-$\alpha$ 
recombination emission from galaxies at 10.2 eV can 
be seen in \citetalias{HM12} but not in our UVB because this emission is 
suppressed in our models due to dust correction. \citetalias{FG09} do not include \hi~Lyman-$\alpha$ 
recombination emission in their calculations. A sharp saw-tooth features due to 
Lyman series absorption of \heii, seen in \citetalias{HM12} UVB at $z>3$, 
are not included in our and
\citetalias{FG09} models. Note that this saw-tooth 
modulation only shows significant difference in UVB at high redshifts and affects a
very small wavelength range.

The differences in the UVB models can be roughly mapped to the differences in their 
predictions of \ghi~and \gheii~(see Fig.~\ref{fig4}). 
The UVB model of the \citetalias{FG09} is significantly different than our and 
\citetalias{HM12} model at all wavelengths and redshifts.
Our UVB model at $z<2$ has higher \hi~ionizing UVB as compared to both \citetalias{FG09} 
and \citetalias{HM12} models. At $z<1$, the similar \heii~ionizing UVB seen for ours and \citetalias{HM12} model is because of the 
coincidental combinations of different $\alpha$ and $\epsilon_{\nu}^Q$ 
used in both models. At higher redshifts our \heii~ionizing UVB
shows larger deviation from \citetalias{HM12}. 
The similarity of X-ray background in our and \citetalias{HM12} at $z=0$
is due to the fact that both models match the X-ray background measurements, 
which was not attempted by \citetalias{FG09}. 
The \heii~ionizing UVB in \citetalias{FG09} shows comparatively smaller 
intensities with larger troughs at $z<2$, suggesting that
\heii~is recombining quickly in their model. This is not surprising given 
the small recombination time scales of the \heii, the ever 
decreasing \gheii~ at $z<2$ and their small values at higher $z>2$ (see right-hand panel of Fig.~\ref{fig4}). 
Sources of all these significant differences among the three UVB models 
are the differences in the input parameters used in modeling, mainly the 
emissivities from QSOs and galaxies along with the \ghi~measurements that 
are used to check or calibrate the models.

The significant differences in the UVB models highlight the need for
routinely updating the models using ever improving measurements of important input parameters.
It is also important that the photoionization calculations which depend extensively on the assumed UVB model 
should use updated UVB keeping in mind all the uncertainties involved in the calculations of UVB. 
This is the main reason
we provide six other UVB models having different $\alpha$ values than our 
fiducial model. For these models, the differences in the obtained 
UVB spectrum can be seen from Fig.~\ref{figA4} of Appendix. The intensity of 
\heii~ionizing UVB increases with $\alpha$ at all $z$
since it depends only on QSO emissivity, on the other hand the \hi~ionizing 
UVB depends on both QSOs and galaxies therefore it only shows
clear dependences on $\alpha$ for $z<3$ where QSOs are dominating the UVB. 
As mentioned earlier, the intensity of \heii~ionizing UVB
is more sensitive to $\alpha$ than \hi~ionizing UVB, as can be seen from Fig.~\ref{figA4} of Appendix for UVB at $z<3$. 
We make all these UVBs publicly available for testing the dependence of 
photoionization calculation results on the assumed spectral shape  of the 
UVB and to be able to quantify the uncertainties in the inferred quantities from these calculations.  
In the following section we give a detailed account of  
the uncertainties involved in the UVB calculations.

\subsection{Uncertainties and caveats in the UVB models}\label{sec4.4}
The synthesis models of EBL are affected by several assumptions and 
many uncertainties arising from various input parameters.
Here, we discuss such uncertainties and caveats 
in our UV and X-ray background models (${\rm E} >13.6$ eV). We refer interested
readers to Section~9 of \citetalias{Khaire15ebl} for the discussion 
related to uncertainties in the far-UV to FIR parts (${\rm E} < 13.6$ eV) of the EBL.

\subsubsection{QSO emissivity and SED}\label{sec4.4.1}
Uncertainties in QSO emissivity arises from how well we measure the QSO luminosity function
(especially at low luminosities) and how representative is the SED used in the wavelength
range where there are no direct observations.
At $z<2$ the QLFs are relatively well measured having shallow faint end slopes. 
Therefore, uncertainties in the obtained emissivity at $z<2$ by integrating down to
faintest luminosity is small. For example, for $z<2$ QLFs from \citet{Croom09} and \citet{Palanque13},
changing the minimum luminosity
$L_{\nu}^{\rm min}=0.01L^*$ to $L_{\nu}^{\rm min}=0$ changes the emissivity by less than 5\%.
Although we are using most recent measurements, the $z>3$ QLFs at the low-luminosities are not
well-measured which can make our assumed emissivities highly uncertain at these redshifts.

For a fixed type-1 QSO SED, any effect of a change in QSO emissivity will be compensated by a
corresponding change in the $f_{\rm esc}$ to match the \ghi~measurements. However, this 
required change in $f_{\rm esc}$ can affect the spectral shape of the \hi~ionizing UVB depending on 
the differences in the type-1 QSO SED and the intrinsic SED of galaxies at ${\rm E} > 13.6$ eV.
Latter is obtained from stellar population synthesis models. 
For our fiducial UVB model, since the SED of QSOs (with $\alpha=-1.8$) 
is co-incidentally same as our intrinsic SED of galaxies, using different QSO emissivity will not affect even the
spectral shape of \hi~ionizing UVB.  
However, the \heii~ionizing UVB will be affected since it depends on the emission from type-1 QSOs and 
the \ghi~through radiative transfer effects that 
determine $\eta$~\citep[see][]{Khaire17sed}\footnote{The change in $f_{\rm esc}$ can affect the 
\heii~ionizing UVB but only if the \ghi~is allowed to vary \citep{Khaire13}.}.
For example, there is no need to have seemingly unrealistic sharp increase in 
$f_{\rm esc}$ with $z$ (the left-hand panel of Fig.~\ref{fig3}) if 
the QSO emissivity is significantly higher than our fiducial values at $z>3.5$. A much 
slower or no evolution in $f_{\rm esc}(z)$ can be easily achieved as shown in \citet{Khaire16} 
when one uses high QSO emissivity models based on the QLF measurements by \citet{Giallongo15} and assuming
that the low-luminosity QSOs also have high (close to unity) escape fraction \citep[see][]{Grazian18}. 
However, such models have serious problems to reproduce the \heii~optical depth 
as a function of $z$ \citep{Khaire17sed}. Moreover QLF measurements of 
\citet{Giallongo15} are not supported by other similar
studies \citep[see e.g.,][]{Weigel15, Ricci17, McGreer17}.

Even the latest measurements of type-1 QSO SED show large variation in the measured value of 
power-law index $\alpha$ at ${\rm E}>13.6$ eV \citep[from -2.3 to -0.5 within 1-$\sigma$ 
errors;][]{Stevans14, Lusso15, Tilton16}. See table 1 of \citet{Khaire17sed} for a summary 
of $\alpha$ measurements till date. The change in \hi~ionizing UVB arising due to change in 
$\alpha$ can be seen in Fig.~\ref{figA4} for $z\le2$ where the UVB is dominated by only 
type-1 QSOs. At $z>3$, Fig.~\ref{figA4} shows no variation in \hi~ionizing UVB with change 
in $\alpha$ since it is dominated by emission from galaxies. The \heii~ionizing part of the 
UVB, however shows a large variation with $\alpha$ at all redshifts, since it only depends on type-1 
QSO emissivity. The existing measurements of $\alpha$ probe smallest wavelength only upto 
425 \AA~(30 eV), therefore there are no direct observational constraints on QSO SED at 
\heii~ionizing wavelengths. It has been assumed that the type-1 QSO SED at 
$\lambda < 912$ \AA~(E$>$13.6 eV) follows a single power law (with same $\alpha$) which is 
normalized at $912$~\AA~or higher wavelengths.
This extrapolated power-law to \heii~ionizing wavelengths gives large intensity differences corresponding to small changes 
in $\alpha$. Therefore, \heii~ionizing UVB and \gheiiz~are more sensitive to $\alpha$ values than the \hi~ionizing UVB and
\ghiz~as can be observed from Fig.~\ref{figA1} and \ref{figA4} of Appendix. Under the assumption of a single power-law, 
in \citet{Khaire17sed} we find that $\alpha$ can have values from -1.6 to -2.0 
consistent with measurements of \citet{Lusso15}
but smaller than measurements of \citet{Stevans14} and \citet{Tilton16}.  
We provide UVB models with varying $\alpha$ from
-1.4 to -2.0 in Appendix~\ref{app.other_uvbs}. Using different $\alpha$ in the UVB models can affect the inferred 
properties of absorbers in the IGM and CGM, 
such as metallicity \citep{Hussain17, Muzahid17}, density \citep{Hussain17, Phoebe17} and temperatures (Gaikwad et al. in prep.).

Type-2 QSO SED and its normalizations, for different values of $\alpha$ used in type-1 QSO SED (see Table~\ref{t1}), 
are adjusted to be consistent with most of the X-ray and $\gamma$-ray background measurements at $z=0$. 
These models with different $\alpha$ are shown in Fig.~\ref{figA3}. For large values of $\alpha$, such as -1.4 or -1.5, it is
difficult to adjust normalizations to be consistent with 
some of the soft X-ray measurements.  Nevertheless, all models are consistent
with measurements at energies more that 20 keV all the way upto 
TeV. Although, this shows a major success of such type-2 QSO and blazar
SED formulation, it has been assumed to scale with type-1 QSO emissivity, 
with the same scaling at all redshifts. Such scaling can be justified using 
recent observations of QSO population \citep[][]{Lusso13, Georgakakis17, Vito17}, 
where the fraction of type-2 QSOs is shown to be non-evolving with redshift
\citep[however, see][]{Liu17, Gohil17}. Due to this scaling, all the 
uncertainties in type-1 QSO emissivity reflect in the 
normalization of type-2 QSO SED. Although, 
it will not affect the obtained X-ray background it will change the 
interpretation related to the fraction of type-2 QSOs as discussed in the Appendix~\ref{app.fraction}. 
However, one should take such an interpretation with 
caution since the  type-1 and type-2 QSO SEDs at X-ray can be adjusted arbitrarily
to give different fractions of type-2 QSOs while still 
being consistent with local X-ray and $\gamma-$ray background measurements.

The soft-X ray background (0.3 keV $<{\rm E}<$ 2 keV) in our EBL model is contributed only by QSOs. However, as shown in
\citet{Phoebe17}, a significant contribution at $z\sim0$ can come from hot intra-halo gas. The contribution
from interstellar gas, CGM and X-ray binary is relatively small, however it depends on the SFRD used in the models.
\citet{Phoebe17} used SFRD from \citetalias{HM12} which is a factor of $\sim 3$ smaller than other estimates at $z<1$. 
In our EBL models we have not included soft-X ray background arising from these different sources which when included
in EBL models can further constrain the SED of type-1 and 2 QSOs and can also change the interpretation of 
fraction of type-2 QSOs as mentioned above. 

\subsubsection{Galaxy emissivity and escape fraction}\label{sec4.4.2}
The emissivity of galaxies at $\lambda < 912$~\AA~depends, in addition to the
intrinsic emissivity, on the $f_{\rm esc}$. In our
model by construct  $f_{\rm esc}$ is constrained to reproduce the
\hi~ionizing UVB consistent with \ghi~measurements.
The intrinsic emissivity at  $\lambda < 912$~\AA~depends on the derived SFRD and $A_{\rm FUV}$ as well 
as several parameters in the stellar population models. The derived SFRD and $A_{\rm FUV}$ alone can have 
systematic uncertainties of the order of 30\% arising due to scatter in various FUV galaxy
luminosity functions 
\citepalias[see e.g., figures 1 and 5 of][]{Khaire15ebl}.
Uncertainties arising in SFRD and $A_{\rm FUV}$  
from different parameters in stellar population models, such as using different 
metallicity are smaller than this uncertainty (see section 8.2 of \citetalias{Khaire15ebl} for more details).

As long as the intrinsic SED of galaxies at $\lambda <912$~\AA~is
same, the intensity of \hi~ionizing UVB remains same for different inferred  
values of $f_{\rm esc}$ arising from different values of 
SFRD, $A_{\rm FUV}$, IMF, metallicity, or other parameters in the stellar population model. 
This is because in the UVB calculations the intrinsic galaxy SED has been 
scaled with $f_{\rm esc}$ \citepalias[see also][]{HM12, FG09}, in absence of any observational 
constrains on the SED of emitted light from galaxies at $\lambda <912$~\AA. 
Such a scaling is justified under the assumption that these 
photons escape through gas and dust free channels, 
as proposed in many theoretical models \citep[e.g.,][]{Fujita03, Gnedin08, Paardekooper11, Conroy12}. 
Although, the intrinsic SED at $\lambda <912$~\AA, shows reasonably small variations for 
different properties of stellar populations such as metallicity 
\citep[see section 4 of][]{Becker13} and different synthesis models
\citepalias[see section 9 of][]{HM12}, different IMFs can provide quite different intrinsic SEDs.

If independent constraints on  $f_{\rm esc}$ are available then the above mentioned
uncertainties will translate to uncertainties in the UVB spectrum. In our models, for a given QSO emissivity, the 
$f_{\rm esc}$ values are adjusted to match \ghi~measurements in the absence of strong constraints 
on the $f_{\rm esc}$ measurements. In future, independent constraints on $f_{\rm esc}$ or \hi~ionizing emissivity 
will be useful to quantify the contributions of sources other than QSOs and galaxies to the UVB.

%
%
\begin{figure*}
\centering
\includegraphics[totalheight=0.745\textheight, trim=0.0cm 0.1cm 7.5cm 0.0cm, clip=true, angle=270]{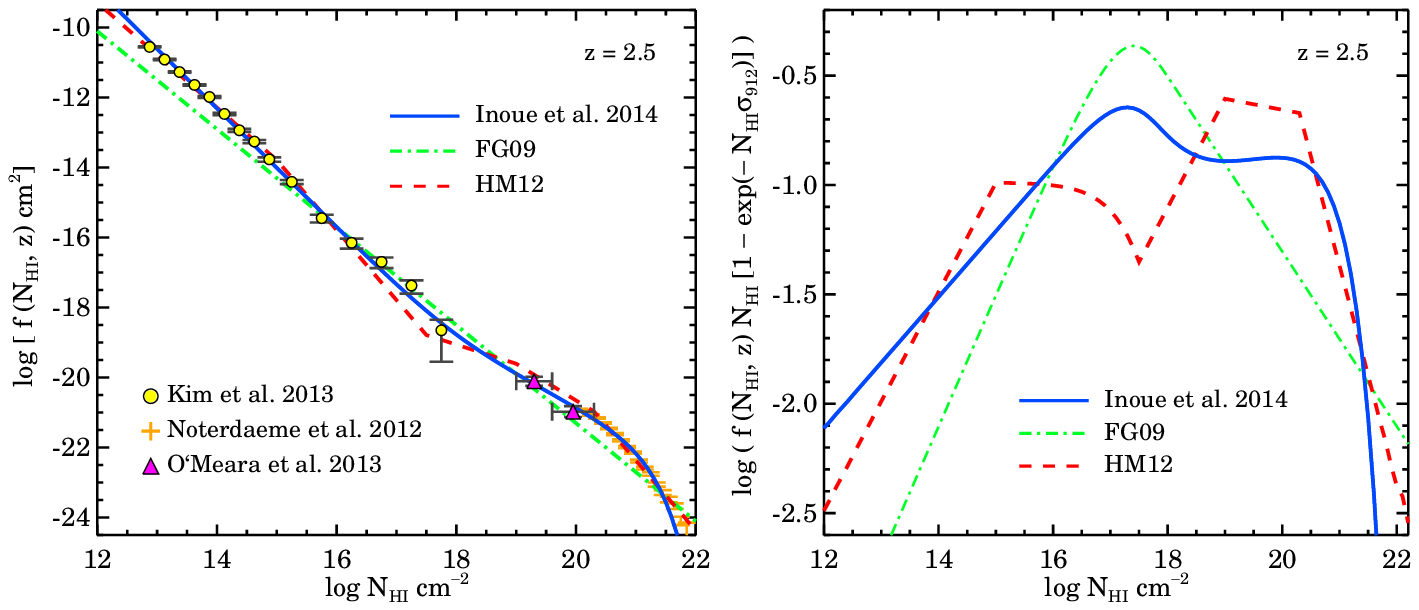}
\caption{ \emph{Left-hand panel}:
The $f(N_{\rm HI}, z)$ at $z=2.5$ from observations at different $N_{\rm HI}$ 
\citep{Kim13, Noterdaeme12, OMeara13} and the fits used by \citetalias{HM12} and \citetalias{FG09} along with 
the updated fits given in \citet{InoueAK14}. \emph{Right-hand panel}: A quantity
$N_{\rm HI}\,f(N_{\rm HI}, z\,) [1-\exp(-N_{\rm HI} \sigma_{0})]$ that determines the 
contribution of $N_{\rm HI}$ to $\tau_{\rm eff}$. For  $f(N_{\rm HI}, z)$ from \citet{InoueAK14}, the column density range
log $N_{\rm HI} (\rm cm^{-2})$
from 15 to 21 (and $4/\eta$ times \nhi) dominates the $\tau_{\rm eff}$ for \hi~(and \heii) ionizing photons.}
\label{fig8}
\end{figure*}
%
\subsubsection{\hi~distribution and other uncertainties}\label{sec4.4.3}
The column density distribution of the \hi, $f(N_{\rm HI}, z)$ plays an important role in shaping the UVB model. 
The $f(N_{\rm HI}, z)$ affects UVB estimate through the calculation of $\tau_{\rm eff}$, especially at 
$15<{\rm log}(N_{\rm HI } \,\, {\rm cm^{-2}})<20.5$ for \hi~ionizing UVB (see Fig.~\ref{fig8}) 
and $4/\eta$ 
times values for \heii~ionizing UVB.
This \nhi~range has relatively higher measurement uncertainty as compared to 
log$(N_{\rm HI } \,\, {\rm cm^{-2}})<14$ or $>20$ due to saturation of the Lyman-$\alpha$ absorption. 
We have used a $f(N_{\rm HI}, z)$ fitting form provided by \citet{InoueAK14} because
it has been obtained by fitting a large number of different observations covering large redshift and \nhi~range, 
as mentioned in Section~\ref{sec2}. For comparison, in the left-hand panel of Fig.~\ref{fig8}, 
we show the $f(N_{\rm HI}, z)$ measurements at $z=2.5$ along with the fitting 
forms from \citet{InoueAK14}, \citetalias{FG09} and \citetalias{HM12}. 
The fit from \citet{InoueAK14} is better than others at all \nhi.  
The difference in the obtained $\tau_{\rm eff}$ at each \nhi~for different  $f(N_{\rm HI}, z)$ fits are 
shown in the right-hand panel of Fig.~\ref{fig8}, where a contribution of $\tau_{\rm eff}$ at each 
\nhi~is represented by quantity $N_{\rm HI} f(N_{\rm HI}, z) [1- exp(-N_{\rm HI} \sigma_{912})]$.
When we use the $f( N_{\rm HI}, z)$ from \citetalias{HM12} instead of \citet{InoueAK14},
at $z<3$ our \ghi~reduces by 10\% (at $z\sim 2.5$ and by 25\% at $z\sim 0.5$) to 40\% (at $z \sim 0$) and \gheii~increases by similar amount  
(because of lower $\eta$ due to small \ghi). These changes are smaller than or comparable to the current measurement 
uncertainties on \gheii~and \ghi~\citep[see figure 16 of][]{Gaikwad17a}. 
We also note that such a change in $f( N_{\rm HI}, z)$ does not affect the 
spectral shape of the UVB significantly \citepalias[see also Fig. 9 of][]{HM12}. 
However, note that, even if we use different $f( N_{\rm HI}, z)$
that can change the UVB, the $f_{\rm esc}$ will be adjusted to get the same \ghi~measurements, 
which further reduces differences in \gheii~as well. 
The change in \ghi~and \gheii~is less than the change arising from varying the $\alpha$ by 0.2. Therefore, 
the uncertainties in $f( N_{\rm HI}, z)$ have smaller impact on the UVB compared to the uncertainties in other parameters,
especially the SED of type-1 QSOs.

The value of $\eta$ calculated using equation~(\ref{quad}) in optically thick \heii~regions 
depends on the assumption that the discrete absorbers in the IGM have line-of-sight thickness equal to the Jeans scale.
Changing this thickness (by factors upto $\sim$ 4) can affect the estimate of \heii~ionizing UVB due to change in $\eta$, 
however its impact on the UVB is quite small since it affects only small number of high column density absorbers.

\subsubsection{Caveats from basic assumptions}\label{sec4.4.4}
Main caveats in the \hi~and \heii~ionizing part of the UVB is that 
the assumption of uniform and homogeneous UVB, which is not valid
at all redshifts. It is certainly a reasonable assumption for 
\hi~ionizing UVB at $z<5.5$ and \heii~ionizing UVB at $z<2.5$
where respective reionization events are believed to be completed. 
The equation~\ref{rad_t} can provide reasonable 
estimates of UVB only in the regime where mean free path of (\hi~and \heii)
ionizing photons is large.
Large fluctuations in the UVB are expected at redshifts where the
respective reionizations are still in progress. At these redshifts, 
the homogeneous \hi~(\heii) ionizing UVB calculated using 
equation~\ref{rad_t} and its predictions for photoionization and 
photoheating rates can not capture the effects of large fluctuations 
in the radiation fields.
As shown by \citet{Jose17}, these rates (from \citetalias{FG09} and \citetalias{HM12})
will prematurely heat the IGM and reionize the \hi~and \heii~very early giving observationally 
inconsistent reionization histories and incorrect feedback on galaxy 
formation. Note that, even for our UVB models the consistency with reionization constraints shown in
Section~\ref{sec4.4.2} (left and middle panel of Fig.~\ref{fig5}), 
is just a sanity check on the ionizing emissivity used in UVB calculations 
and independent of the UVB estimates. Therefore at these redshifts, 
a large scatter expected in the photoheating and photoionization rates should be properly implemented 
in the cosmological hydrodynamic simulations. This can be achieved by modifying these rates
as shown in \citet{Jose17}. Such a modification depends on how the physics of reionization is implemented in 
simulations, the assumed evolution of mean free path for 
ionizing photons and the SED of ionizing sources \citep{Jose17, Puchwein18}. 
Moreover, in absence of the observational constraints on the mean 
free paths and SEDs of ionizing sources at these redshifts, such a 
modification to the photoionization and photoheating rates can not be unique.
Novel methods for determining accurate timing and heat-injection 
during reionization \citep[e.g.,][]{Padmanabhan14, Jose17a} will prove 
valuable in future to resolve these issues.

Even after completion of the reionization, significant fluctuations in 
the UVB are expected \citep[e.g.,][]{Furlanetto09H, Furlanetto09He, Davies16, Davies17}.
In addition, the recently reionized gas 
may take some time to reach ionization and thermal equilibrium.
Our calculations do not consider these non-equilibrium conditions \citep[e.g.,][]{Puchwein15}.
Also, most of the measurements of \ghi~that we used to constrain the ionizing emissivity from galaxies, are
obtained from observed Lyman-$\alpha$ forest by modeling the IGM under the assumption of ionization and thermal 
equilibrium. A consolidated efforts are needed to get these issues sorted-out around the epoch of \hi~and 
\heii~reionization. 

In the UVB calculations, several assumptions related to the input 
parameters are made in absence of the concrete observational constraints. For 
example, the QSO SED has been assumed to be the same at all redshifts 
in absence of significant observational evidence of its redshift 
evolution \citep[][]{Stevans14, Tilton16}. Also, it has been 
assumed that the type-1 QSO SED follows a single power-law for 
$\lambda<912$~\AA~upto wavelengths $\sim 50$ \AA, although available
observations only probe upto $425$~\AA. Moreover, a single power-law may not
be a good approximation to small wavelengths \citep[e.g,][]{Tilton16, Khaire17sed}. 
Similarly for galaxies, although mild metallicity evolution does
not affect intrinsic SEDs, the IMF and other stellar population 
parameters have been assumed to be the same at all redshifts.  The obtained $f_{\rm esc}$
and SED of galaxies can be different for top heavy IMFs 
\citep{Topping15} and including stellar rotation \citep{Ma15} and binary stars 
\citep{Rosdahl18} in the population synthesis models.
For galaxies, a crucial assumption, as mentioned before, 
is the scaling of intrinsic galaxy SED by $f_{\rm esc}$.  
Because of this our another assumption, a single dust extinction 
curve at all $z$, does not affect UVB but
can change EBL at FIR wavelength due to the corresponding change in $A_{\rm FUV}$ and SFRD 
that preserves the emissivities till near-IR wavelengths \citepalias[see][]{Khaire15ebl}.

In spite of all these uncertainties, we provide EBL models that are consistent  
with currently available observations at all redshifts and wavelengths. Better constraints on
\ghi~at 0.5$\le z\le$2.0 and at $z>5$, well defined QLFs at low luminosity end and more
measurements of $\tau_{\rm HeII}$ at $z>2$ will provide better constraints on the UVB models.

One of the important implications of the UVB for IGM is its thermal evolution \citep{Jose17, Puchwein18}. 
We find that our photoheating rates when used in hydrodynamical simulations of the IGM performed with explicit
non-equilibrium ionization effects provide IGM thermal histories consistent with most of the recent measurements
(Gaikwad et al. in prep.).

\section{Summary}\label{sec5}

The EBL is extensively used by the astronomical 
community for studying (i) spectral energy distribution
of blazars in GeV energies, (ii) metal line absorption 
systems to derive density, metallicity and size of
the absorbing gas and their redshift evolution and (iii) 
ionization and thermal state of the IGM over cosmic time.
As the spectrum of EBL can not be directly measured at all epochs, 
it has to be modeled using the available 
observations of the source emissivities and IGM opacities 
with  appropriate cosmological radiative transport. 
Thus it is important to have EBL computed time to time with 
latest parameters that govern the source emissivities 
and IGM opacities. In addition, it is important to quantify 
the allowed variations in computed EBL at each epoch so that
one can estimate systemic uncertainties arising from EBL 
uncertainties while interpreting the data.

With these motivations, we present new synthesis models of 
the EBL which cover more than fifteen orders of magnitude 
in wavelength from FIR to TeV $\gamma$-rays (see 
Fig.~\ref{fig6} for $z=0$ EBL). Our main focus of this paper is 
on modeling the observationally consistent extreme-UV and 
soft X-ray background, which is essential 
for studying the metal absorption lines observed in QSO spectra. 

In our EBL synthesis model, we use updated inputs such 
as SFRD and $A_{\rm FUV}$ for galaxies, QSO emissivity
and \hi~distribution of the IGM. We determine average 
escape fraction ($f_{\rm esc}$) of \hi~ionizing photons from galaxies which
can reproduce the available recent measurements of the 
\hi~photoionization rates at $z<6.5$ (the left-hand panel of 
Fig.~\ref{fig3} and Fig.~\ref{fig4}) and provide 
\hi~reionization history consistent with many new measurements 
of \hi~fraction at $z>6$ and electron scattering optical depth to CMB (Fig.~\ref{fig5}). 
The UV background at $z<3$ is predominantly contributed by
QSOs whereas galaxies contribute significantly to the \hi~ionizing 
emissivity at $z>3$ \citep[see also][]{Khaire16, Gaikwad17a}
as evident from a sharp increase in the required $f_{\rm esc}(z)$ (see Fig.~\ref{fig3}). 

For our fiducial EBL model, we take type-1 QSO extreme 
UV SED as $f_{\nu} \propto {\nu}^{\alpha}$
with power-law slope $\alpha=-1.8$ from \citet{Khaire17sed}, 
which consistently reproduce the $2.5<z<3.5$
measurements of Lyman-$\alpha$ effective optical depths, photoionization rates
and reionization redshift of \heii~(the right-hand panel of Fig.~\ref{fig4} and Fig.~\ref{fig5}).
Since available measurements of $\alpha$ show large variation, we also provide six other EBL models 
(in Appendix) calculated for different values of $\alpha$ from -1.4 to -2.0 
consistent with recent measurements of $\alpha$ from \citet{Shull12}, \citet{Stevans14} and \citet{Lusso15}. 
At X-ray and $\gamma$-ray energies, we modified the form of 
type-1 QSO SED and constructed a SED for type-2 QSOs 
(including $\gamma$-ray emitting blazars) following 
\citet{Sazonov04} to consistently reproduce the 
measurements of local X-ray and $\gamma$-ray background 
(Fig.~\ref{figA3}). We find that, EBL models constructed in this way
with $\alpha > -1.5$ are not consistent with \heii~Lyman-$\alpha$ effective optical depths
\citep[see also][]{Khaire17sed} and some of the soft-X-ray  background measurements.

Because the UV background is an important tool 
to study the time evolution in physical and chemical properties of the IGM and CGM,
it needs to be consistent with various observational constraints. We discuss
in details the uncertainties and caveats arising from 
several assumptions in the modeling of UV background (see Section~\ref{sec4.4}). 
We are working on addressing some of the caveats such as non-equilibrium ionization
and thermal evolution of the IGM and its impact on the UV background around the epochs of \hi~and 
\heii~reionization \citep[see e.g,][]{Gaikwad18}.
Within the framework discussed here and in the previous UV background models, 
a major uncertainty lies in the spectral shape, therefore improved observational constraints on the QSO SEDs
and escaping Lyman-continuum SEDs of galaxies are important. 
Also, new techniques to constrain spectral shape of the UV background by using metal absorption lines
are required \citep[e.g.,][]{Fechner11,Finlator16}. Our UV background models obtained for different QSO SEDs
will be useful for such studies.

For public use we provide the EBL tables, the photoionization
and photoheating rates of \hi, \hei~and \heii~obtained from EBL models, 
and $\gamma$-ray effective optical depths at all redshifts.\footnote{All tables are available at 
\href {http://www.iucaa.in/projects/cobra/} {http://www.iucaa.in/projects/cobra/} and 
\href {http://vikramkhaire.weebly.com/downloads.html} {http://vikramkhaire.weebly.com/downloads.html}}

\section*{acknowledgement} 
We thank Marco Ajello, Nico Cappelluti and Eugene Churazov for providing the X-ray background data.
We thank Prakash Gaikwad, Tirthankar Roy Choudhury, Aseem Paranjape and Kandaswamy Subramanian for helpful 
discussions throughout the development of this project. VK thanks Fred Davies, 
Jose Onorbe, Tobias Schmidt and other \href {http://enigma.physics.ucsb.edu/}{ENIGMA} group members at UCSB 
for valuable comments on the manuscript.  

\bibliographystyle{mnras}
\bibliography{vikrambib}

\appendix
\section{Escape fraction measurements}\label{app.escape}
In Table~\ref{tab.escape}, we provide the details of the average 
escape fraction measurements which are shown in the left-hand panel
of the Fig.~\ref{fig3}. The $f_{\rm esc}$ is obtained using following relation
\begin{equation}\label{eq.B1}
 f_{\rm esc}=f^{\rm rel}_{\rm esc}10^{-0.4A_{\rm FUV}}\Big(\frac{f_{\lambda_{\rm FUV}}}{{f_{\lambda_i}}}\Big)_{\rm KS15}
\Big(\frac{f_{\lambda_{\rm FUV}}}{{f_{\lambda_i}}}\Big)^{-1}
\end{equation}
where, $f^{\rm rel}_{\rm esc}$ is the relative escape fraction 
provided in the respective references assuming an intrinsic value
of $(f_{\lambda_{\rm FUV}}/f_{\lambda_i}$) where $\lambda_{\rm FUV}$ denotes the rest wavelength at FUV and
the $\lambda_{i}$  denotes rest wavelength where ionizing photons have been observed. 
The $(f_{\lambda_{\rm FUV}}/f_{\lambda_i})_{\rm KS15}$ denotes the intrinsic value of 
$(f_{\lambda_{\rm FUV}}/f_{\lambda_i}$) and $10^{-0.4A_{\rm FUV}}$ is the dust correction
from our fiducial galaxy emissivity model. We use the 1-$\sigma$ value of $f_{\rm esc}$ given in 
\citet{Matthee17} without any correction since
they use an indirect method to obtain $f_{\rm esc}$ from H$\alpha$ emission.
 
\begin{table*}
\caption{Escape fraction measurements.}
\def\arraystretch{1.5}
\begin{tabular}{ l c c c c c c c c} 
\hline
(1)       &(2)        &(3)             &(4)                 &(5)                                                  &(6)                                   &(7)                                                                 &(8)                    &(9)           \\      
Reference & $z$-range & $z_{\rm mean}$ & $\lambda_i$ (\AA)  & $\Big(\frac{f_{\lambda_{\rm FUV}}}{f_{\lambda_i}}\Big)$  & $f^{\rm rel}_{\rm esc}$ (1-$\sigma$) & $\Big(\frac{f_{\lambda_{\rm FUV}}}{f_{\lambda_i}}\Big)_{\rm KS15}$ & $10^{0.4A_{\rm FUV}}$ & $f_{\rm esc}$ (1-$\sigma$) \\
\hline
\citet{Rutkowski16}   & $0.90-1.40$ &1.15 & 900 & 7.0  & $< 0.027$        & 5.63 & 6.07 & $<0.0036$ \\
\citet{Siana10}       & $1.20-1.50$ &1.35 & 700 & 3.0  & $< 0.006$        & 8.40 & 5.91 & $<0.0028$ \\
\citet{Rutkowski17}   & $2.38-2.90$ &2.56 & 900 & 3.0  & $< 0.07$         & 5.26 & 3.88 & $<0.0316$ \\
\citet{Micheva17}$^a$ & $3.06-3.13$ &3.1  & 900 & 4.25 & $< 0.228$        & 5.06 & 3.2  & $<0.0849$ \\
\citet{Micheva17}$^b$ & $3.06-3.13$ &3.1  & 900 & 4.25 & $< 0.06$         & 5.06 & 3.2  & $<0.0223$ \\
\citet{Grazian17}$^c$ & $3.27-3.40$ &3.3  & 900 & 3.0  & $< 0.103$        & 5.0  & 3.0  & $<0.0572$ \\
\citet{Guaita16}      & $3.11-3.53$ &3.4  & 900 & 5.0  & $< 0.12$         & 5.18 & 2.91 & $<0.0427$ \\
\citet{Japelj17}$^d$  & $3.0-4.0$   &3.5  & 900 & 3.0  & $< 0.20$         & 4.96 & 2.83 & $<0.1168$ \\
\citet{Marchi17}      & $3.5-4.5$   &3.81 & 895 & 3.0  & 0.09$\pm$0.04& 5.6  & 2.6  & 0.065$\pm 0.029$ \\
\citet{Matthee17}$^e$& $2.20-2.24$&2.22 & H$\alpha$ &                &  &  & & $<0.028$ \\
\hline
\end{tabular}
\begin{flushleft}
\footnotesize{{\bf Notes:}} \footnotesize{Corresponding to the references 
provided in column(1), the column(2) and (3) provides range in redshifts and the 
mean redshift for the observed sample of galaxies, column (4) 
provides rest frame extreme-UV wavelength $\lambda_i$ where Lyman-continuum
has been observed, column (5) gives the fiducial values of  
$f_{\lambda_{\rm FUV}}/f_{\lambda_i}$ and column (6) gives the 1-$\sigma$ values of 
relative escape fraction $f^{\rm rel}_{\rm esc}$ obtained in these references. Column (6) and (7) 
provides fiducial values of  $f_{\lambda_{\rm FUV}}/f_{\lambda_i}$
denoted as $(f_{\lambda_{\rm FUV}}/f_{\lambda_i})_{\rm KS15}$ and the dust corrections ($10^{0.4A_{\rm FUV}}$) 
used in our galaxy emissivity model. Column (7) provides the absolute 
escape fraction values relevant to our galaxy model obtained by
using equation~\ref{eq.B1}. These are plotted in the left-hand panel of Fig.~\ref{fig3}.} \\
\footnotesize{$^a$ Observed for a sample of Lyman-$\alpha$ emitters.}\\
\footnotesize{$^b$ Observed for a sample of Lyman-break galaxies.}\\
\footnotesize{$^c$ Obtained for the galaxies with luminosity $>0.2$ L$^*$.}\\
\footnotesize{$^d$ Obtained for the galaxies with luminosity $>0.5$ L$^*$.}\\
\footnotesize{$^e$ The measurement is obtained using the H$\alpha$ emission from galaxies.}
\end{flushleft}
\label{tab.escape}
\end{table*}


\section{X-ray emissivity with parametric SED}\label{x-ray_emissivity}
%
\begin{figure*}
\centering
\includegraphics[totalheight=0.38\textheight, trim=0.0cm 0.0cm 0.0cm 0.0cm, clip=true]{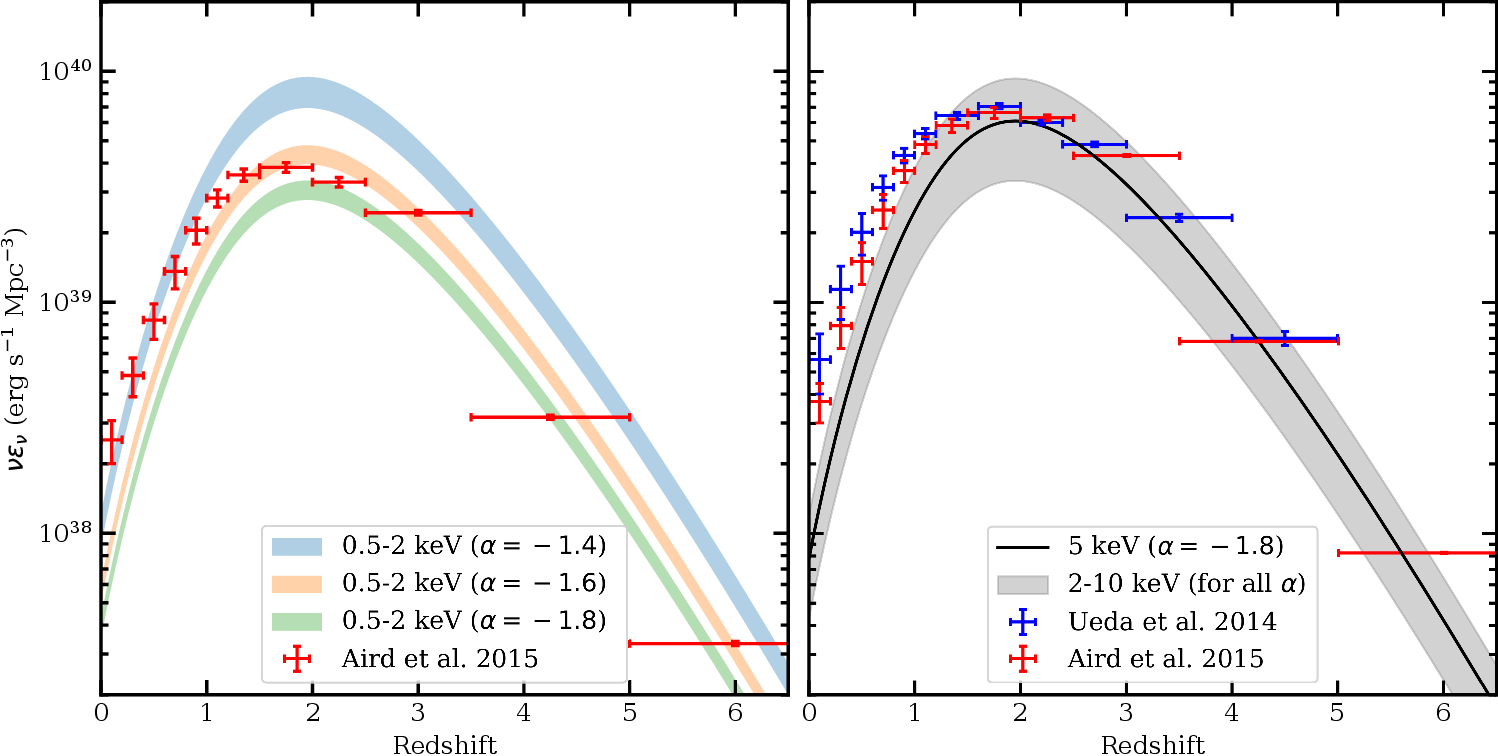}
\caption{
Comparison of the X-ray emissivity obtained from X-ray luminosity functions and predicted by our parametric SED formulation
as described in section~\ref{sec3.1}.
Left panel shows soft X-ray emissivity (0.5-2 keV range) and right-hand panel shows
hard X-ray emissivity (2-10 keV range). 
The emissivities obtained from our parametric approach are shown with colored shaded regions. The data points 
are obtained from the X-ray luminosity functions given in \citet{Ueda14} and \citet{Aird15} using their LDDE models. 
Vertical error-bars denote the range in emissivity calculated by setting the minimum luminosity in the range of  $L_{\nu_0}^{\rm min}= 0.01 - 0.1 L_{\nu_0}^*$. 
Horizontal error-bars demarcate the redshift bin of the luminosity function. In soft X-ray regime our emissivity depends on the choice of $\alpha$, therefore
we show predictions for three values of $\alpha$. At $z<1$, measurements are consistent with $\alpha=-1.4$, while at $z>1$ they are more in line with $\alpha < -1.6$. 
For hard X-ray regime our emissivity is independent of choice of $\alpha$  and  at $z<1$ our predictions are systematically lower than the measurements. 
In both cases, our model predictions follow measured redshift evolution of X-ray emissivity reasonably well.
} 
\label{figAA}
\end{figure*}
%
To obtain the X-ray emissivity, we used a simple parametric approach to construct X-ray SED of QSOs (see 
section~\ref{sec3.1}) following \citet{Sazonov04} than full X-ray population synthesis. Even though our simple
approach consistently reproduces the local EBL measurements at X-ray and $\gamma$-ray energies, it is important to
validate its predictions at different redshifts. Therefore, in this section we compare the redshift evolution of
our predicted X-ray emissivity with the measurements. 

We use recent X-ray luminosity functions in soft ($0.5-2$ keV) and hard ($2-10$ keV) X-ray band
by \citet{Ueda14} and \citet{Aird15} and calculate the emissivity  
with $L_{\nu_0}^{\rm min}=0.1-0.01 L_{\nu_0}^*$ in equation~(\ref{eq.rho}). We used their luminosity functions parameters 
obtained for luminosity dependent density evolution (LDDE) 
case.  Note that these emissivities are obtained from the intrinsic 
luminosity functions and therefore they can be considered as upper limits for the comparison with our model. 
The redshift evolution of this intrinsic emissivity along with our model
predictions are shown in Fig.~\ref{figAA}. 

Left-hand panel of Fig.~\ref{figAA} shows soft X-ray emissivity along with our predictions for different values of
$\alpha$ where shaded regions enclose the relevant energy range of $0.5-2$ keV. At these energies our predictions depend on the choice of $\alpha$. 
At $z<1$ the measurements are consistent with $\alpha=-1.4$ and higher than the predictions for $\alpha <-1.4$, whereas
at $z>1$ they are more in line with $\alpha < -1.6$. Right-hand panel of Fig.~\ref{figAA} shows hard 
X-ray emissivity along with our prediction for $\alpha=-1.8$ for relevant energy range of $2-10$ keV. At these energies our 
predictions are independent of the choice of $\alpha$. At $z<1$ our predictions are systematically lower than the measurements but
consistent at higher-$z$. Nevertheless, the redshift evolution of the emissivities in both soft and hard X-ray regime is reproduced quite well by
our models. Note that our parametric SED does not evolve with redshift but the normalization which is anchored with respect to type-1 QSO 
emissivity evolves. The reasonable agreement among our predictions and the intrinsic emissivities and their redshift evolution, along with its success to
reproduce the local X-ray and $\gamma$-ray background justify our simple parametric approach.

\section{$\gamma$-ray opacity}\label{app.gamma_tau}
%
%
\begin{figure}
\centering
\includegraphics[totalheight=0.30\textheight, trim=0.6cm 0.1cm 2.2cm 0cm, clip=true]{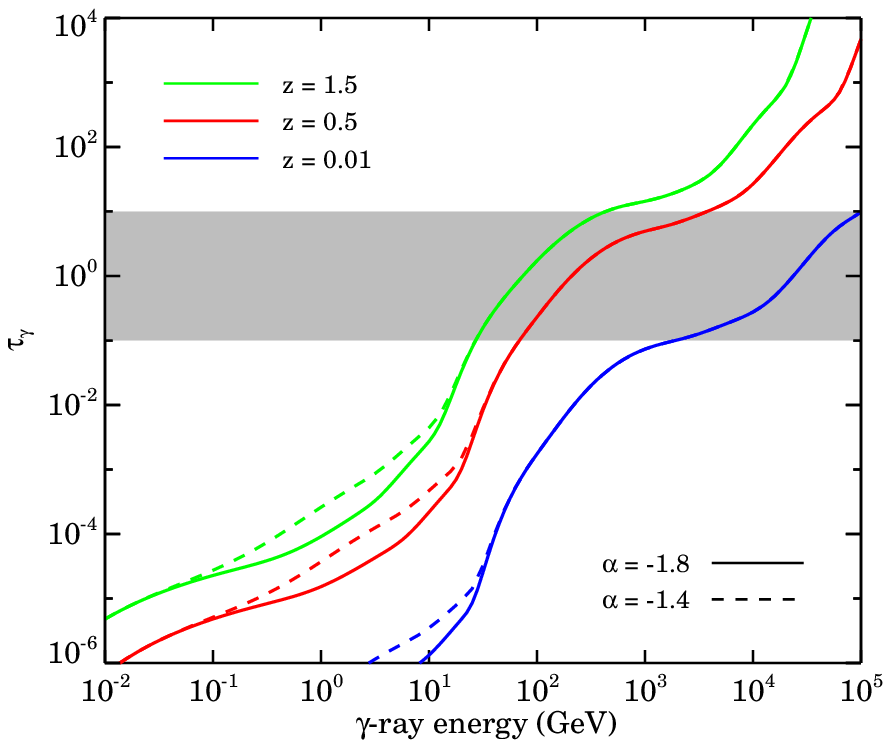}
\caption{Optical depth ($\tau_{\gamma}$) encountered by $\gamma$-rays 
for different energy (in GeV) observed on Earth which were emitted from
redshift $z=0.01$ (blue curves), $z=0.5$ (red curves) and $z=1.5$ (green curves). 
Solid and dashed curves are for our EBL models with $\alpha=-1.8$ (fiducial)
and $\alpha=-1.4$. $\tau_{\gamma}$ is calculated by including CMB in both EBL models.
Shaded region shows observationally relevant  $\tau_{\gamma}$ values from 0.1 to 10 which
are insensitive to the EBL photons with energy $\rm E<10 $ eV and thus on values of assumed $\alpha$.}
\label{figA6}
\end{figure}
%
%
The optical depth encountered by $\gamma$-rays depends on the energy of the gamma-rays, their source redshift $z_0$ and
the number density of the EBL photons at each $z\le z_0$. If $n(E_{ebl},z)$ is the number density of the EBL photons at 
redshift $z$ having energy $E_{ebl}$ per unit energy then the optical depth encountered by $\gamma$-rays
emitted at redshift $z_0$ with frequency $\nu_{\gamma} (1+z_0)$ and observed on Earth at frequency $\nu_{\gamma}$ is given by
\begin{eqnarray}
\lefteqn{\tau_{\gamma}(\nu_{\gamma},z_0) =  \frac{1}{2}\int^{z_0}_0 dz\;\frac{dl}{dz}\int^1_{-1}d(\cos\theta) \; (1-\cos\theta)} \nonumber \\ 
& & \times \int^{\infty}_{E_{min}} dE_{ebl}\; n(E_{ebl},z)\;\sigma(E_\gamma (1+z),E_{ebl},\theta).
\label{Eq.tau}
\end{eqnarray}
Here $E_{\gamma}= h\nu_{\gamma}$, 
\begin{equation}
E_{min}=\frac{2m_e^2c^4}{h\nu_{\gamma}(1+z)(1-\cos\theta)}\,\,,
\end{equation}
and $\sigma(E_\gamma (1+z),E_{ebl},\theta)$ is the cross-section 
for pair production (see equation 14 of \citetalias{Khaire15ebl}).  

Using above equation, we calculate the $\tau_{\gamma}(\nu_{\gamma},z)$ at 
different $z$ for our fiducial EBL model (Q18) and EBL model 
with $\alpha=-1.4$. Note that to estimate $\tau_{\gamma}$ we include CMB in our EBL models. 
Results for three redshifts are shown in Fig.~\ref{figA6}. 
The different features seen in the $\tau_{\gamma}(\nu_{\gamma})$ 
can be naively mapped to the features seen in the EBL. 
For example, the $\gamma$-rays with $E_{\gamma} < 20/(1+z)$ GeV ($0.1/(1+z)$ GeV) are attenuated by EBL 
with $E_{ebl}>13.6$ eV (500 eV), the  $\gamma$-rays with 
$E_{\gamma} > 100/(1+z)$ TeV are affected by CMB, and the bump in $\tau_{\gamma}$ around 100 GeV
is due to rest of the EBL. 
The shaded region in Fig.~\ref{figA6} shows the observationally relevant opacities
from 0.1 to 10. Within this region, EBL models with different $\alpha$ give same $\tau_{\gamma}$ 
since it is dominated by low-energy photons 
($E_{ebl} < 10$ eV) which do not depend on the value of $\alpha$.
Fig.~\ref{figA6} shows that the intrinsic SED of blazars at $E>100$ GeV in our formalism (Section~\ref{sec3}) 
will be severely different that the one given by equation~\ref{sed2}. 
The $\tau_{\gamma}$ values for our fiducial EBL model are available online.$^5$ 


\section{Fraction of type-2 QSOs }\label{app.fraction}
%
\begin{figure*}
\centering
\includegraphics[totalheight=0.745\textheight, trim=0.0cm 0.1cm 7.5cm 0.0cm, clip=true, angle=270]{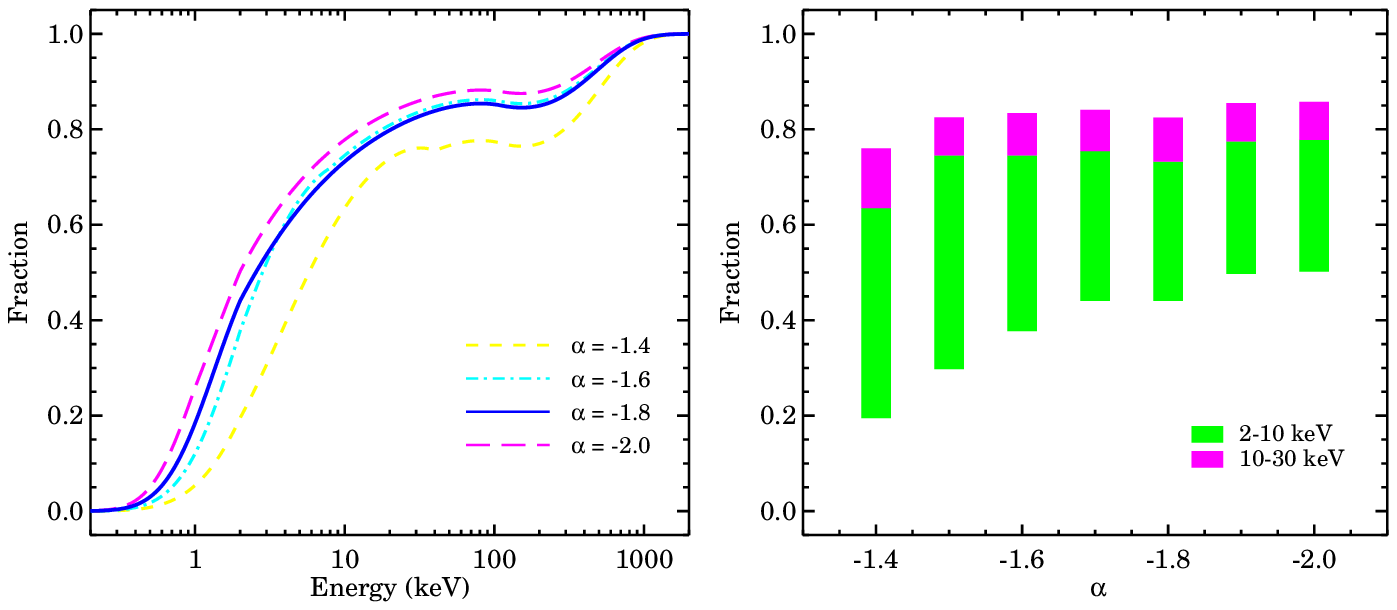}
\caption{
\emph{Left-hand panel}:
The fraction of type-2 QSOs defined as a ratio $k^{\rm Q2}_{\nu}/k_{\nu}$ 
(see Section~\ref{sec3}, equation~\ref{sed} to \ref{sed2})
with energy in keV. The fraction also gives contribution of type-2 QSOs and blazars to EBL relative to type-1 QSOs
for the SED templates used in our model. It depends on the extreme-UV power-law index 
$\alpha$ of the type-1 QSO SED. At very high energies, fraction reaches unity due to blazar 
contribution included in the $k^{\rm Q2}_{\nu}$.
\emph{Right-hand panel}:
The range in the fraction of type-2 QSOs in two energy ranges, 2-10 keV 
and 10-30 keV, for different values of $\alpha$.  The fraction
in the energy range 2-10 keV and 10-30 keV can be thought as the 
fraction of type-2 QSOs that are Compton thin and Compton thick, 
respectively. Out of total type-2 QSOs, 20-30\% contribution comes from Compton thick QSOs. 
}
\label{figA0}
\end{figure*}

In Section~\ref{sec3}, the full SED of QSOs ($k_{\nu}$) has been 
constructed to reproduce the X-ray and $\gamma$-ray background measurements. 
For such a constructed SED, the frequency dependent 
ratio of Type-2 QSO SED and full SED, $k^{\rm Q2}_{\nu}/k_{\nu}$, 
can be interpreted as the fraction of the 
type-2 QSOs which is luminosity weighted and redshift independent. 
This fraction is decided by our constructed 
SED of type-1 QSOs in the X-ray energies, 
which depends on the value of $\alpha$. Note 
that we have also included blazar contribution in $k^{\rm Q2}_{\nu}$.
In the left-hand panel of Fig.~\ref{figA0} we plot 
the ratio $k^{\rm Q2}_{\nu}/k_{\nu}$ for different values of $\alpha$
(from equation \ref{sed} to \ref{sed2}). 
The $\alpha$
dependence can be easily seen in the plot. The small value of 
$\alpha$ leads to less emissivity from type-1 QSOs requiring more contribution 
from type-2 QSOs giving rise to the higher value of their fraction. 
At energies less than 1 keV (2 keV), the fraction of type-2 QSOs is less 
than 0.2 (0.5) for all $\alpha$. At energies less than 500 eV, 
the contribution from type-2 QSOs drops below 5\%.
Type-2 QSOs contribute significantly at hard X-ray energies around 
10-100 keV consistent with the requirement of large number
of highly obscured type-2 QSOs in X-ray population synthesis models. 
At energies more than 1000 keV, the fraction of type-2 QSOs is unity 
owing to the fact that blazars are sole contributers at 
these energies \citep[see also][]{Draper09}. 
For comparison with the definition of the Compton thin and 
Compton thick type-2 QSOs, as the contribution to X-ray background at energies 2-10 keV
and 10-30 keV respectively, we plot a range in the fraction at 
these energies for different values of $\alpha$ 
in the right-hand panel of the Fig.~\ref{figA0}. We find that the 
fraction of Compton thin QSOs to vary between 0.2-0.8 for different $\alpha$. 
It is 0.44-0.73 for our fiducial $\alpha=-1.8$ model. For $\alpha \le -1.5$, it varies from 0.38 to 0.78. 
These values are consistent with recent measurements \citep[for e.g.,][]{Buchner15, Georgakakis17, Vito17} 
as well as with the 
prediction from X-ray population synthesis models \citep[][]{Treister06, Gilli07}. 
We need 20-30\% more contribution in 10-30 keV energy range 
from Compton thick QSOs, consistent with the predictions
from \citet{Draper09}, however, it depends on the contribution of blazars which we do not resolve. 
All this interpretation is subject to our construction 
of type-1 QSO SED in X-rays. One can always adjust it to get different numbers while being
consistent with the observed X-ray and $\gamma$-ray background. Note that, our 
motivation is not to understand the fraction of type-2 QSOs but to provide
an observationally consistent high-energy background.


\section{EBL models with different values of $\alpha$}\label{app.other_uvbs}
\begin{figure*}
\centering
\includegraphics[totalheight=0.745\textheight, trim= 7.5cm 0.0cm 0.1cm 0.0cm , clip=true, angle=90]{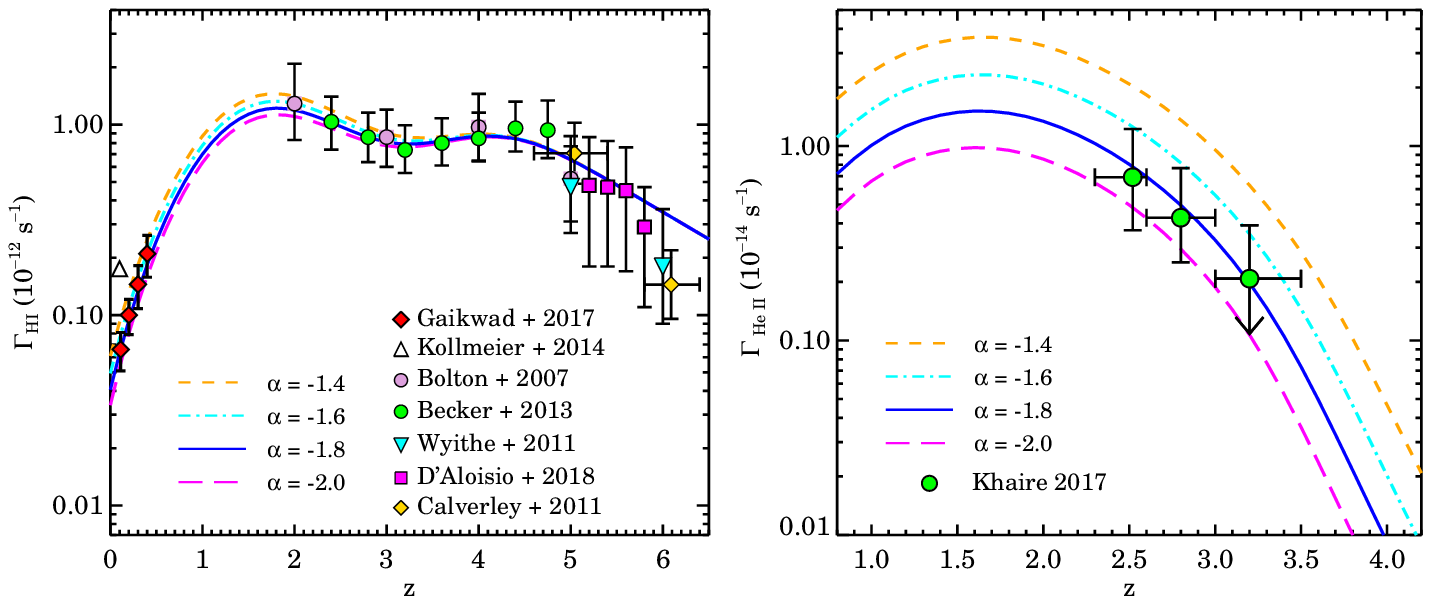}
\caption{
The photoionization rates of \hi~($\Gamma_{\rm HI}$; left-hand panel) 
and \heii~($\Gamma_{\rm He II}$; right-hand panel) with $z$ 
from UVB models with different $\alpha$. 
Various data points in the left-hand panel show the recent measurements of $\Gamma_{\rm HI}$. 
The $\Gamma_{\rm He II}$ from \citet{Khaire17sed} is obtained by using the measurements of 
$\tau_{\alpha}^{\rm He II}$ from \citet{Worseck16}.
}
\label{figA1}
\end{figure*}

\begin{figure*}
\centering
\includegraphics[totalheight=0.745\textheight, trim=0.0cm 0.1cm 7.5cm 0.0cm, clip=true, angle=270]{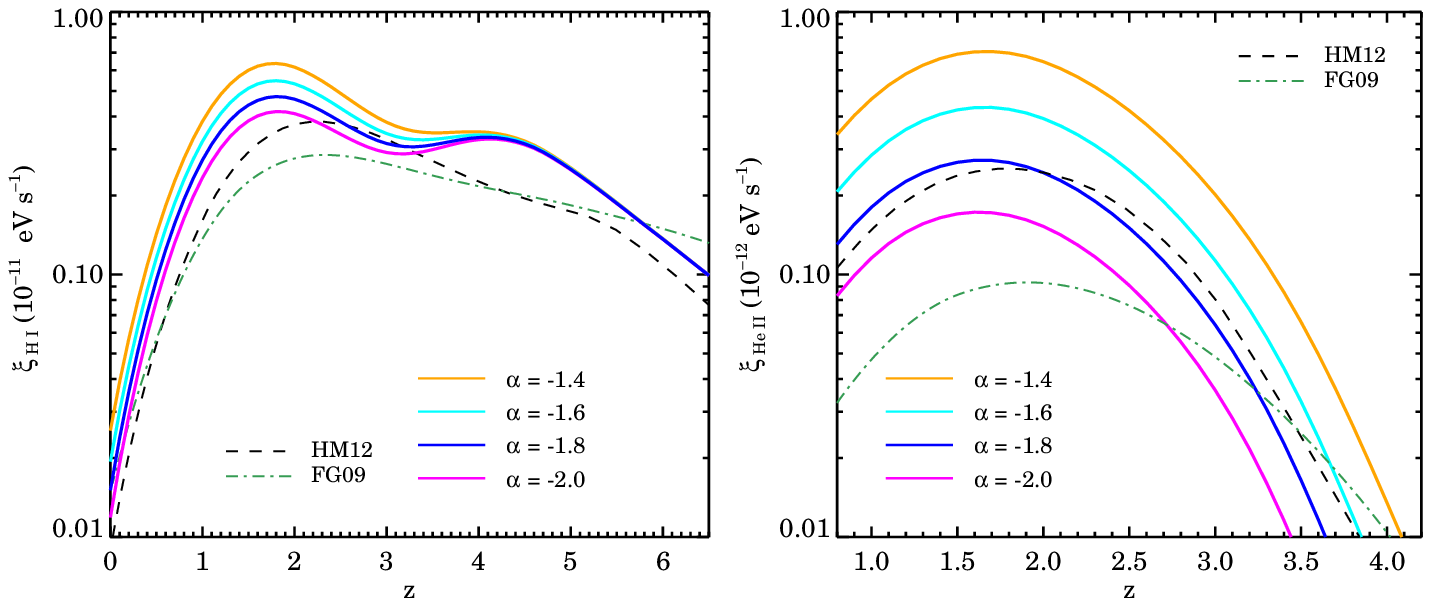}
\caption{
The photoheating rates of \hi~($\xi_{\rm HI}$; left-hand panel) and \heii~($\xi_{\rm He II}$; right-hand panel) with $z$ 
from UVB models with different $\alpha$. The dash and dot-dash curve show the results from 
UVB models of \citetalias{HM12} and \citetalias{FG09}, respectively.
}
\label{figA2}
\end{figure*}

\begin{figure*}
\centering
\includegraphics[totalheight=0.745\textheight, trim=0.0cm 0.1cm 5.2cm 0.0cm, clip=true, angle=270]{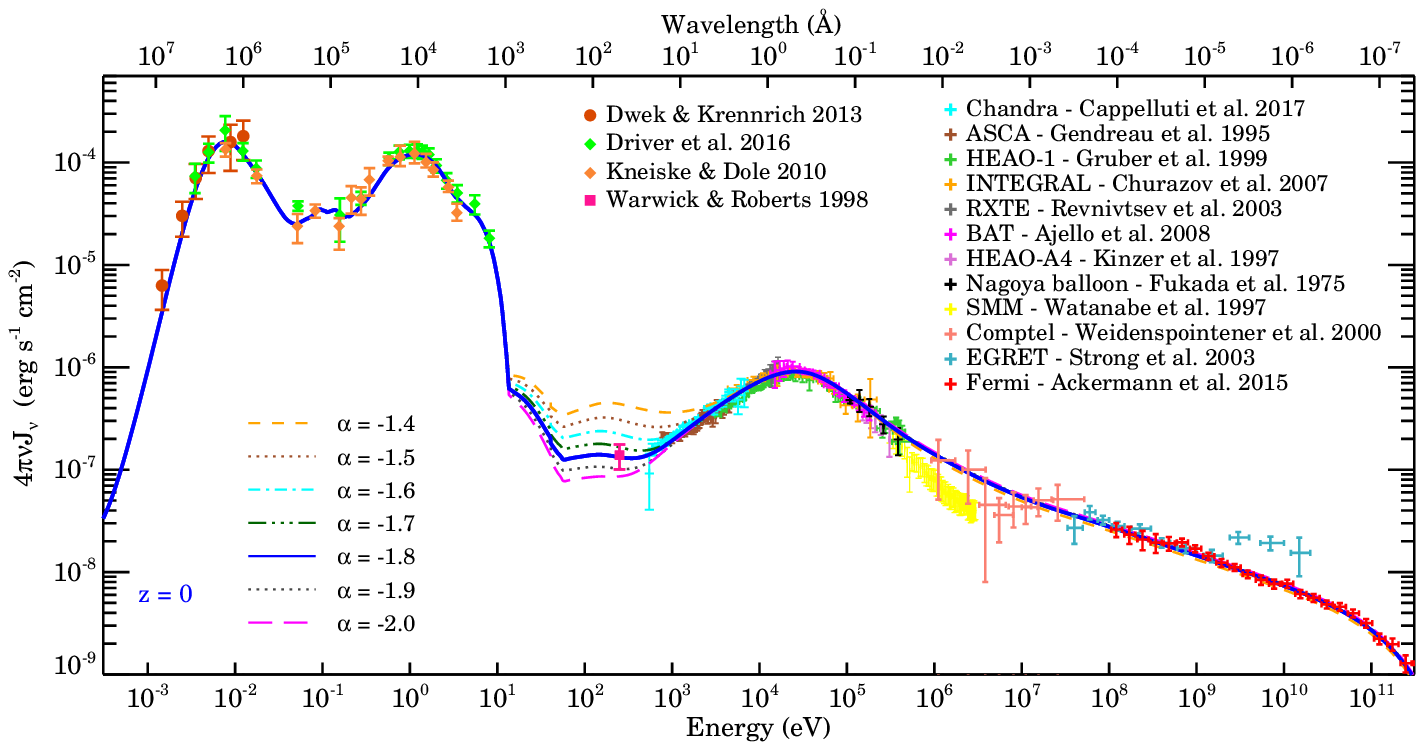}
\caption{The intensity of $z=0$ EBL models with different $\alpha$ from FIR to TeV $\gamma$-rays. 
For details on data points refer to Section~\ref{sec4.3}.
}
\label{figA3}
\end{figure*}

\begin{figure*}
\centering
\includegraphics[totalheight=0.745\textheight, trim=0.0cm 0.1cm 2.8cm 0.0cm, clip=true, angle=270]{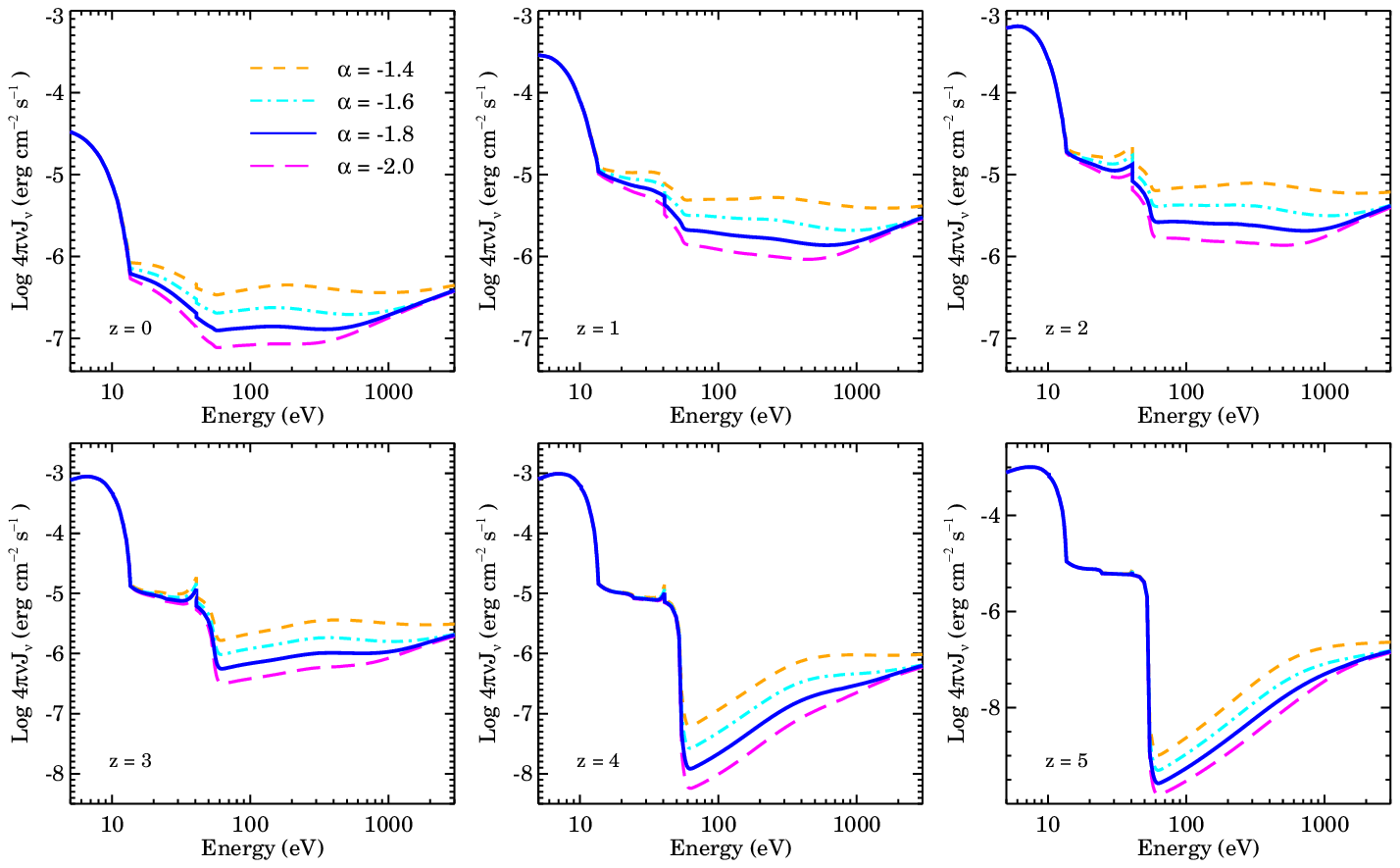}
\caption{Spectrum of UVB  models obtained for different values of $\alpha$ with energy (from 5 to 3000 eV) at $z=0$, 1, 2, 3, 4 and 5. 
Solid blue curves show our fiducial UVB models with $\alpha=-1.8$. The $y$-axis scales are kept same for top three panels and bottom two panels to see relative change in intensity.}
\label{figA4}
\end{figure*}

Here, we provide the EBL models (plots and tables) obtained 
for different values of $\alpha$ ranging from -1.4 \citep{Shull12, Stevans14} 
to -2.0 \citep{Lusso13, Khaire17sed} in the interval of 0.1. 
Fig.~\ref{figA1} shows photoionization rates \ghiz~and \gheiiz, 
Fig.~\ref{figA2} shows photoheating rates $\xi_{\rm HI}$ and 
$\xi_{\rm He II}$ and Tables~\ref{ta2} to \ref{ta8} provide these values for 
EBL models with different $\alpha$ from z=0 to 15 including for \hei. 
Fig.~\ref{figA3} shows the local EBL from FIR to $\gamma$-rays 
and Fig.~\ref{figA4} shows UVB at redshifts $z=0$ to 5. Refer to Section~\ref{sec4} 
for the relevant discussion on the trends seen in these plots. 

%
%
%
%

%
%
%
%
%
%
%

%
\begin{table*}
\caption{Photoionization and photoheating rates for our fiducial Q18 UVB model.}
\begin{tabular}{ c c c c c c c c c c c c c c } 
\hline
 $z$   & \ghi    & \ghei    & \gheii   & \hhi        & \hhei       & \hheii    \\
    & s$^{-1}$ & s$^{-1}$ & s$^{-1}$ & eV s$^{-1}$ & eV s$^{-1}$ & eV s$^{-1}$\\
\hline      
 0.0  &4.058e-14 &1.869e-14 &6.262e-16 & 1.507e-13 & 1.494e-13 & 1.248e-14 \\
 0.1  &6.310e-14 &3.041e-14 &9.453e-16 & 2.381e-13 & 2.383e-13 & 1.857e-14 \\
 0.2  &9.353e-14 &4.695e-14 &1.378e-15 & 3.575e-13 & 3.640e-13 & 2.670e-14 \\
\hline
\end{tabular}
\label{ta2}
\end{table*}
%

\begin{table*}
\caption{Photoionization and photoheating rates for Q14 UVB model.}
\begin{tabular}{ c c c c c c c c c c c c c c } 
\hline
 $z$   & \ghi    & \ghei    & \gheii   & \hhi        & \hhei       & \hheii    \\
    & s$^{-1}$ & s$^{-1}$ & s$^{-1}$ & eV s$^{-1}$ & eV s$^{-1}$ & eV s$^{-1}$\\
\hline  
 0.0   &   6.105e-14   &   3.709e-14  &    1.755e-15  &    2.544e-13   &   3.467e-13  &    3.675e-14 \\
 0.1   &   9.188e-14   &   5.816e-14  &    2.588e-15  &    3.895e-13   &   5.325e-13  &    5.372e-14 \\
 0.2   &   1.324e-13   &   8.706e-14  &    3.692e-15  &    5.691e-13   &   7.863e-13  &    7.588e-14\\
\hline
\end{tabular}
\label{ta3}
\end{table*}

\begin{table*}
\caption{Photoionization and photoheating rates for Q15 UVB model.}
\begin{tabular}{ c c c c c c c c c c c c c c } 
\hline
 $z$   & \ghi    & \ghei    & \gheii   & \hhi        & \hhei       & \hheii    \\
    & s$^{-1}$ & s$^{-1}$ & s$^{-1}$ & eV s$^{-1}$ & eV s$^{-1}$ & eV s$^{-1}$\\
\hline  
 0.0 &5.480e-14      &3.099e-14      &1.339e-15      &2.214e-13      &2.773e-13      &2.745e-14 \\
 0.1 &8.320e-14      &4.907e-14      &1.989e-15      &3.418e-13      &4.304e-13      &4.039e-14 \\
 0.2 &1.208e-13      &7.403e-14      &2.856e-15      &5.031e-13      &6.412e-13      &5.741e-14 \\
\hline
\end{tabular}
\label{ta4}
\end{table*}

\begin{table*}
\caption{Photoionization and photoheating rates for Q16 UVB model.}
\begin{tabular}{ c c c c c c c c c c c c c c } 
\hline
 $z$   & \ghi    & \ghei    & \gheii   & \hhi        & \hhei       & \hheii    \\
    & s$^{-1}$ & s$^{-1}$ & s$^{-1}$ & eV s$^{-1}$ & eV s$^{-1}$ & eV s$^{-1}$\\
\hline  
%
 0.0 &4.944e-14      &2.602e-14      &1.027e-15      &1.938e-13      &2.234e-13      &2.071e-14 \\
 0.1 &7.568e-14      &4.162e-14      &1.536e-15      &3.016e-13      &3.502e-13      &3.064e-14 \\
 0.2 &1.107e-13      &6.333e-14      &2.220e-15      &4.472e-13      &5.266e-13      &4.379e-14 \\
\hline
\end{tabular}
\label{ta5}
\end{table*}

\begin{table*}
\caption{Photoionization and photoheating rates for Q17 UVB model.}
\begin{tabular}{ c c c c c c c c c c c c c c } 
\hline
 $z$   & \ghi    & \ghei    & \gheii   & \hhi        & \hhei       & \hheii    \\
    & s$^{-1}$ & s$^{-1}$ & s$^{-1}$ & eV s$^{-1}$ & eV s$^{-1}$ & eV s$^{-1}$&  \\
\hline  
%
 0.0  &4.468e-14      &2.194e-14      &7.955e-16      &1.701e-13      &1.814e-13      &1.586e-14 \\
 0.1  &6.895e-14      &3.540e-14      &1.196e-15      &2.670e-13      &2.870e-13      &2.357e-14 \\
 0.2  &1.015e-13      &5.425e-14      &1.738e-15      &3.984e-13      &4.350e-13      &3.382e-14 \\
\hline
\end{tabular}
\label{ta6}
\end{table*}

\begin{table*}
\caption{Photoionization and photoheating rates for Q19 UVB model.}
\begin{tabular}{ c c c c c c c c c c c c c c } 
\hline
 $z$   & \ghi    & \ghei    & \gheii   & \hhi        & \hhei       & \hheii    \\
    & s$^{-1}$ & s$^{-1}$ & s$^{-1}$ & eV s$^{-1}$ & eV s$^{-1}$ & eV s$^{-1}$\\
\hline  
%
 0.0  &3.685e-14      &1.586e-14      &4.860e-16      &1.332e-13     & 1.221e-13      &9.640e-15 \\
 0.1  &5.775e-14      &2.604e-14      &7.368e-16      &2.121e-13     & 1.967e-13      &1.439e-14 \\
 0.2  &8.619e-14      &4.048e-14      &1.079e-15      &3.205e-13     & 3.029e-13      &2.075e-14 \\
\hline
\end{tabular}
\label{ta7}
\end{table*}

\begin{table*}
\caption{Photoionization and photoheating rates for Q20 UVB model.}
\begin{tabular}{ c c c c c c c c c c c c c c } 
\hline
 $z$   & \ghi    & \ghei    & \gheii   & \hhi        & \hhei       & \hheii   \\
    & s$^{-1}$ & s$^{-1}$ & s$^{-1}$ & eV s$^{-1}$ & eV s$^{-1}$ & eV s$^{-1}$ \\
\hline  
%
 0.0  &3.364e-14      &1.361e-14      &3.860e-16      &1.187e-13     & 1.015e-13      &7.721e-15 \\
 0.1  &5.309e-14      &2.250e-14      &5.867e-16      &1.903e-13     & 1.649e-13      &1.152e-14 \\
 0.2  &7.974e-14      &3.519e-14      &8.619e-16      &2.893e-13     & 2.555e-13      &1.662e-14 \\
\hline
\end{tabular}
\label{ta8}
\begin{flushleft}
Full Tables~\ref{ta2} - \ref{ta8} in machine-readable format 
are available to download at \href {ftp://ftp.iucaa.in/in.coming/KS18EBL/} {IUCAA-ftp} and a
\href {http://vikramkhaire.weebly.com/downloads.html} {webpage}.\\
\end{flushleft}
\end{table*}

%
\end{document}